\documentclass[usenatbib]{mn2e}
\usepackage{amsmath}
\usepackage{amssymb}
\usepackage[pdftex]{xcolor}
\usepackage{graphicx}

\pdfoutput=1

\newcommand{\Msun}{\,\mathrm{M}_{\odot}}
\newcommand{\Lsun}{\,\mathrm{L}_{\odot}}
\newcommand{\pc}{\,\mathrm{pc}}
\newcommand{\kpc}{\,\mathrm{kpc}}
\newcommand{\yr}{\,\mathrm{yr}}
\newcommand{\Myr}{\,\mathrm{Myr}}
\newcommand{\Gyr}{\,\mathrm{Gyr}}
\newcommand{\kms}{\,\mathrm{kms}^{-1}}

\newcommand{\Nbody}{$N$-body }
\newcommand{\RG}{R_\mathrm{G}}
\newcommand{\VG}{V_\mathrm{G}}
\newcommand{\Exp}{\mathrm{exp}}
\newcommand{\sigeta}{\sigma_{(\eta)}}

\title[Dynamical Constraints on the Origin of MSPs in GCs]{Dynamical Constraints on the Origin of Multiple Stellar Populations in Globular Clusters}

\author[P. Khalaj and H. Baumgardt]{P. Khalaj\thanks{E-mail: pouria.khalaj@uqconnect.edu.au} and H. Baumgardt\\
School of Mathematics and Physics, University of Queensland, St. Lucia, QLD 4072, Australia}

\begin{document}

\date{Accepted xxxx. Received xxxx; in original form xxxx}

\pagerange{\pageref{firstpage}--\pageref{lastpage}} \pubyear{2015}
\maketitle
\label{firstpage}

\begin{abstract}
We have carried out a large grid of \Nbody simulations in order to investigate if mass-loss as a result of primordial gas expulsion can be responsible for the large fraction of second generation stars in globular clusters (GCs) with multiple stellar populations (MSPs). Our clusters start with two stellar populations in which 10\% of all stars are second generation stars. We simulate clusters with different initial masses, different ratios of the half-mass radius of first to second generation stars, different primordial gas fractions and Galactic tidal fields with varying strength. We then let our clusters undergo primordial gas-loss and obtain their final properties such as mass, half-mass radius and the fraction of second generation stars. Using our \Nbody grid we then perform a Monte Carlo analysis to constrain the initial masses, radii and required gas expulsion time-scales of GCs with MSPs. Our results can explain the present-day properties of GCs only if (1) a substantial amount of gas was present in the clusters after the formation of second generation stars and (2) gas expulsion time-scales were extremely short ($\lesssim10^5\yr$). Such short gas expulsion time-scales are in agreement with recent predictions that dark remnants have ejected the primordial gas from globular clusters, and pose a potential problem for the AGB scenario. In addition, our results predict a strong anti-correlation between the number ratio of second-generation stars in GCs and the present-day mass of GCs. So far, the observational data show only a significantly weaker anti-correlation, if any at all.
\end{abstract}

\begin{keywords}
globular clusters: general -- stars: chemically peculiar -- stars: formation -- stars: kinematics and dynamics -- methods: numerical
\end{keywords}


\section{Introduction}\label{sec:introduction}
It is generally assumed that all stars in star clusters are born in close proximity to each other, and in well-mixed molecular clouds by a rapid star formation process and therefore have similar ages and metallicities \citep{Lada}. As a result, star clusters should only host a single population of stars, i.e. they are bona fide {\bf single stellar population} systems. However, recent observations of GCs show a statistically significant star-to-star variation in the abundance of light elements, such as Na, O, Mg or Al (e.g., \citealt{Carretta09, Gratton13}). These abundance anomalies of light elements are not associated with any spread in the iron abundance for the majority of GCs except for a few cases such as $\omega$ Cen \citep{Gratton04}. Such massive GCs are thought to be different than normal GCs and have a different origin, for example being the remnant of a disrupted dwarf galaxy \citep{Meza}. 

\par In addition to the abundance anomalies mentioned above, the colour-magnitude diagrams of some GCs split into two or more evolutionary sequences (e.g., $\omega$ Cen \citealt{Rey} and \citealt{Bedin}; NGC 2808 \citealt{DAntona04}, \citealt{Piotto} and \citealt{Milone12c}; NGC 1851 \citealt{Milone08}; 47 Tuc \citealt{Milone12a}; NGC 6397 \citealt{Milone12b}; M22 \citealt{Marino}; GCs in Fornax \citealt{DAntona13}). 

\par These findings are indicative of self-enrichment in GCs and suggest that star clusters are comprised of at least two stellar populations, in direct contradiction to the conventional star formation scenario described earlier. We refer to these two populations as first generation (FG) and second generation (SG) stars for convenience and to be consistent with previous studies such as \citet{Decressin08}. In our terminology FG and SG stars correspond to stars with normal (or primordial) and enriched chemical compositions respectively.

The observations show that the number ratio of SG to FG stars, $N_2/N_1$, is around unity although with considerable spread \citep{DAntona08}. 
Further studies on MSPs have shown that they exhibit different spatial ($\omega$ Cen \citealt{Bellini}; several GCs \citealt{Lardo}; 47 Tuc \citealt{Nataf}; M15 \citealt{Larsen15}) and dynamical signatures (47 Tuc \citealt{Richer} and \citealt{Kuvcinskas}). However, \citet{Dalessandro} found that the different populations in NGC 6362 share the same radial distribution which is the first evidence of fully spatially mixed MSPs ever observed in a GC. 

\par Several scenarios have been proposed to address the origin of multiple stellar generations in GCs (\citealt{Decressin07a, Decressin07b, deMink, Renzini, DErcole10, Conroy, Ventura01, Valcarce} and \citealt{Bastian13}) among which the four main scenarios are: (1) fast rotating massive stars (FRMS) ($20-120\Msun$; \citealt{Prantzos, Maeder, Decressin07a, Decressin07b}); (2) asymptotic giant branch (AGB) stars ($4-9\Msun$; \citealt{Ventura01, DErcole08, DErcole10, Ventura11}); (3) massive ($10-100\Msun$) and intermediate mass ($4-10\Msun$) binaries \citep{deMink} and (4) early disc accretion in low-mass pre-main-sequence stars (enriched gas comes from stars with $M>10\Msun$)\citep{Bastian13}. 

\par In the FRMS scenario, stars that spin at a rate close to their critical break-up speed can lose extensive amounts of mass via stellar winds which are slow enough to be retained in the gravitational potential well of the cluster and form circumstellar discs out of which SG stars will be born. In the AGB and massive binary scenarios, these slow winds come from the envelopes of evolving intermediate-mass stars in their AGB phase and numerous massive interacting binaries in the core of clusters respectively. SG stars in the FRMS scenario need to form in a very short time-scale, $t<8.8$\,Myr \cite{Krause13}, before the burst of the first supernova (SN), since SN winds can destroy circumstellar discs formed around fast-rotating FG stars and interrupt the star formation \citep{Decressin07a}. In the AGB scenario, on the other hand, the formation of SG stars is triggered after all SNe have gone off and the cluster has been cleared of SN II ejecta ($t>28$\,Myr) \citep{DErcole08}, otherwise AGB ejecta will be polluted by SN ejecta which will cause the iron abundance of SG stars to differ from that of FG stars, in contradiction with observations. \citet{Bastian13} proposed a model in which GCs do not need to go through different instances of star formation to produce chemically peculiar stars. According to this model, interacting massive binaries ($M>10\Msun$) supply the intra-cluster medium with enriched material which will be accreted by low-mass stars ($M<2\Msun$) while they are still in their pre-main-sequence phase. The main difference between this model and other models is that stars with different chemical abundances belong to the same generation of stars. The main caveat of this model is that the circumstellar discs around accreting low-mass stars need to survive for 5 to 10\,Myr which is a questionable assumption  in GCs with a denser core \citep{Bastian13}. 

\par In a recent study, \citet{Bastian15} tested the yields of all the proposed models in the literature for their consistency with observations and concluded that none of the models is able to explain the observed He abundance of clusters. As a result the origin of abundance anomalies in GCs is still a matter of debate. 

\par In addition to this discrepancy between the theoretical yields and observations, the ejecta in the FRMS and AGB scenarios are not enough to form a large population of SG stars and explain the roughly equal number of FG and SG stars found in observations \citep{DAntona08}. Assuming a canonical \citet{Kroupa} initial mass function (IMF), the total mass which is lost by all FRMS and AGB stars, constitutes between $\sim 4\%$ to $9\%$ of the initial mass of all FG stars \citep{deMink}. If we assume that the gas which is lost by this mechanism entirely turns into SG stars, i.e. star formation efficiency is $100\%$, the number ratio of SG stars to FG stars is expected to be at almost $\sim 10\%$. This issue, which is referred to as the mass-budget problem, does not exist for the massive and intermediate-mass binaries as they provide more ejecta than AGB and fast-rotating massive stars combined. In addition their ejecta are further mixed with an approximately equal amount of pristine gas which doubles the mass of the available gas for star formation. As a result there is a substantial amount of polluted gas to form a large number of SG stars \citep{deMink}. 

To address the mass-budget problem two solutions have been proposed: either a cluster must have been at least 10-20 times more massive and have undergone significant mass-loss ($\sim 90\%$) or the cluster IMF must have been strongly top-heavy, i.e. it initially had many more massive stars than predicted by a canonical IMF. 

There is observational evidence against both of these solutions. First, observations of GCs in a number of dwarf galaxies show a high ratio of metal-poor GCs to field stars which cannot be explained if star clusters were initially 10 times more massive and underwent significant mass-loss \citep{Larsen12, Larsen14}. Second, \citet{Dabringhausen} found that a top-heavy IMF will lead to high mass-to-light ($M/L$) ratios in old stellar systems ($t=12$\,Gyr) such as UCDs and GCs. This is a serious issue for the AGB scenario, as the polluters in the AGB scenario evolve into white dwarfs and the retention factor of white dwarfs is very high compared to the FRMS scenario in which polluters evolve into black holes or neutron stars, many of which will leave the cluster. In the AGB scenario, depending on whether SG stars form as a distinct generation or they are only contaminated by the processed gas during formation, one needs a high-mass slope of $\alpha=-1.15$ and $\alpha=-1.95$ respectively\footnote{The high-mass slope of a \citet{Kroupa} IMF is $\alpha=-2.35$ } to provide enough ejecta from AGB stars to form low-mass stars (Scenarios I and II of \citealt{Prantzos}). According to Fig. 2 of \citet{Dabringhausen}, this will translate into a normalised $M/L$ ratio of $5.3$ and $4.2\Msun\Lsun^{-1}$ if the retention factor of SN remnants is 0 and $6.5$ and $4.3\Msun\Lsun^{-1}$ if it is 20\%. The average observed $M/L$ ratio of GCs in the Milky Way, its satellites and M31 is less than $2.0\Msun\Lsun^{-1}$ \citep{McLaughlin, Strader}. If one considers the biases that exist in the derivation of masses from integrated light of GCs, the observed $M/L$ ratios can be explained by a canonical IMF \citep{Shanahan}. 

\par The focus of the present paper is to study the effect of significant mass-loss on the dynamical evolution of star clusters with MSPs. The dynamical effects of a top-heavy IMF on GCs with MSPs can be further examined in a future paper. Using \Nbody simulations we determine the required initial conditions under which the final number of SG to FG stars match the observations. We then perform a Monte Carlo (MC) analysis to compare the outcome of our simulations with observations and determine whether the significant mass-loss scenario is able to explain the observed mass and half-mass radius distributions of GCs and ultimately be the reason for the observed abundance anomalies.

\par The present paper is structured as follows. In the next section we briefly review the mass-loss mechanisms which can affect the dynamical evolution of GCs. We discuss the details of our \Nbody simulations and the procedures we have followed to create our grid of runs in Section \ref{sec:nbodySimulations}. In Section \ref{sec:results} we present our results and then compare the outcome of simulations with observations using a MC analysis in Section \ref{sec:observations}. We finally conclude our work in Section \ref{sec:conclusions}.


\section{Mass-Loss Mechanisms}
There are several mechanisms through which a star cluster can lose mass: stellar evolution induced mass-loss, two-body relaxation, external tidal shocks and primordial gas loss. These mechanisms are discussed further below.

\subsection{Stellar evolution induced mass-loss}
To lose a large amount of mass via stellar evolution, clusters need to be very extended and in a strong tidal field where the ratio of the tidal radius to the half-mass radius, $r_t/r_h$, is small. The effect of stellar evolution induced mass-loss in the AGB scenario has been studied by \citet{DErcole08} using a series of \Nbody simulations. In their models, they use \citet{King} models with $M_{\rm FG}=10^7\Msun$, $r_t=200\pc$ and $W_0=7.0$, $c=\log(r_t/r_c)=1.50$, where SG stars are highly concentrated in the innermost regions of the clusters with a half-mass radius one-tenth of that of the initial FG stars. Such initial parameters correspond to very extended clusters ($r_t/r_h=8.75$ and $r_h=23\pc$) in which SG and FG stars are dynamically decoupled from each other. As a result FG  stars can readily expand their orbits and be stripped by the Galactic tidal field in response to a small amount of mass-loss, leaving a sub-cluster of SG stars in the center of the initial cluster. \citet{DErcole08} also study other models with different initial truncation radii and concentrations which underfill their Roche lobes, but the only models that match the observed number ratio of SG to FG stars, i.e. $N_2/N_1\sim1.0$, are the very extended and tidally filling clusters. This is not in agreement with the observation of young massive star clusters and today's properties of GCs as they have typical half-mass radii of around $1.0\pc$ \citep{PortegiesZwart10} and $5.0\pc$ respectively \citep{Harris}. Unless the condition under which GCs have formed were significantly different, the stellar evolution induced mass-loss cannot lead to significant mass-loss. In addition, it is unclear if a sample of clusters with these initial conditions can explain the observed distribution of SG number ratios in GCs.

\subsection{Two-body relaxation}\label{subsec:twobody}
\citet{Baumgardt03} studied the effect of two-body relaxation as well as stellar evolution on the dynamical evolution of star clusters in external tidal fields through \Nbody simulations. They assumed different Galactic orbits, stellar density profiles and particle numbers for star clusters and derive the following formula for the lifetime of a star cluster
\begin{equation*}
	\frac{T_{\rm diss}}{\rm Myr} = \beta\left[\frac{N}{\ln(0.02N)}\right]^x\frac{\RG}\kpc\left(\frac{\VG}{\rm 220\kms}\right)^{-1}(1-\epsilon)
\end{equation*}
where $N$ is the number of particles, $\VG$ and $\RG$ are the Galactic circular velocity and distance of the cluster and $\epsilon$ is the eccentricity of the cluster orbit. $x$ and $\beta$ are two parameters whose values depend on the initial concentration of the cluster and for King $W_0=7.0$ they are equal to 0.82 and 1.03 respectively.

\par For $N=10^6$, $\epsilon=0.5$, $\VG=220\kms$ and $\RG=8.5\kpc$, $T_{\rm diss}$ will be about $55$\,Gyr which shows that two-body relaxation is a slow process for massive GCs and is not efficient in reducing the mass of GCs by $90\%$ over one Hubble time. As a result this process cannot be the origin of significant mass-loss in star clusters and we will omit this process in our \Nbody simulations. 

\par For a cluster whose initial number ratio of SG to FG stars is $\sim10\%$ and SG stars are more concentrated than FG stars, two-body relaxation causes different stellar populations to fully mix in about 2 elapsed half-mass relaxation times \citep{Decressin08}. Using Eq. 1 of their paper, a cluster with $M=10^6\Msun$ and $r_h=3\pc$, has a mixing time of approximately $2\Gyr$. This implies that any significant mass-loss scenario proposed to explain the origin of MSPs must have a shorter time-scale, since after the mixing has occurred the number ratio of SG to FG stars will not change due to further mass loss.

\subsection{External tidal shocks}
External tidal shocks such as encounters with giant molecular clouds (GMCs) are able to disrupt open clusters $M\leq10^4\Msun$ via a single encounter on time-scales of about $\sim2.0\Gyr$ \citep{Wielen, Gieles06}. \citet{Gieles06} derived the following formula for the disruption time of star clusters
\begin{equation*}
	T_{\rm dis}=2.0\left(\frac{5.1\Msun^2\pc^{-5}}{\Sigma_n\rho_n}\right)\left(\frac{M_c}{10^4\Msun}\right)^{0.61}\Gyr
\end{equation*}
where $\Sigma_n$ and $\rho_n$ are the individual surface and global density of the GMCs, equal to $170\Msun\pc^{-2}$ and $0.03\Msun\pc^{-3}$ in the solar neighbourhood \citep{Solomon}. For a GC with $M_c=10^6\Msun$, this formula gives a disruption time of almost $\sim33\Gyr$. In denser environments such as the center of M51, $\rho_n$ is 10 times higher \citep{Gieles06} which shortens the disruption time by an order of magnitude, but this is still larger than the mixing time of MSPs ($\sim 2\Gyr$), as discussed in Section \ref{subsec:twobody}. In addition encounters with GMCs are stochastic by nature. Hence they cannot be responsible for significant mass-loss in all clusters and their effect is insignificant over short time-scales.

\subsection{Primordial gas loss}
If the star formation efficiency is less than 100\%, this process will happen to every cluster of any size or mass since it has an intrinsic origin. Any gas loss in GCs will be accompanied by loss of stars, especially when the gas loss is impulsive \citep{Baumgardt07}. There are a number of different sources which can inject enough energy into the intra-cluster medium to entirely unbind the primordial gas. Examples are stellar winds, SN explosions \citep{Decressin10} and black holes \citep{Krause12,Krause13}.

\par With all other mass-loss scenarios excluded or shown to be ineffective on short time-scales, primordial gas loss remains as the only plausible and universal mechanism via which GCs can lose a significant amount of mass over a few Myrs and in our \Nbody simulations we only deal with such primordial gas loss as discussed in the next section. As mentioned above accretion onto dark remnants is one candidate for a mechanism which can cause such a primordial gas loss. The setup of our model clusters and our analysis, though consistent with the dark remnant scenario (Section \ref{sec:observations}), is not limited to this scenario and in principle can be applied to any other physical process that has a similar effect on GCs.


\begin{table*}
	\centering
 	\caption{Initial conditions of the simulations.}
	\label{tab:initialConditions}
	\begin{tabular}{lcc}
		\hline
		\hline
		Parameter  & Symbol & Value \\   
		\hline
		{\bf Fixed Parameters} & & \\ \\ 
		Total number of stars & $N$ & 20480 \\ \\
		Initial number of SG to FG stars & $\displaystyle\frac{N_2}{N_1}$ & $0.1$ \\ \\
		Ratio of the Plummer scale radius of the gas cloud to SG stars & $\displaystyle\frac{a_g}{a_2}$ &  $1.0$ \\ \\ \\
		{\bf Variable Parameters} & & \\ \\ 
		Ratio of the Plummer scale radius of SG stars to FG stars & $\lambda=\displaystyle\frac{a_2}{a_1}$ &  $\{0.1, 0.2\}$ \\ \\
		Ratio of the Initial mass of the gas cloud to the mass of FG stars  & $\eta=\displaystyle\frac{M_g}{M_1}$ &  $0.0\leq\eta\leq2.0$ \\ \\   
		Ratio of the gas expulsion time-scale to the initial crossing time  & $\tau=\displaystyle\frac{T_\Exp}{T_{\rm cr}}$ &  $10^{-2}\leq\tau\leq10^4$  \\ \\ 
		Ratio of the initial tidal radius to the Plummer radius of FG stars & $\displaystyle\frac{r_t}{a_1}$ & $\{5, 10, 15, 20, 25, 30, 35, 40, \infty\}$  \\
		\hline 
		\hline
	\end{tabular}	
\end{table*}

\section{\emph{N}-body simulations}\label{sec:nbodySimulations}
We set up clusters consisting of 3 components: FG stars ($\sim90\%$ of total stellar mass), SG stars ($\sim10\%$) and a gas cloud whose mass is a free parameter in the simulations. The initial number ratio of SG to FG stars $N_2/N_1$ is fixed at $0.1$. We do not directly simulate the gas particles but only calculate the force that the gas cloud exerts on each star. The initial density profiles of the different components are given by \citet{Plummer} models with different masses and Plummer radii. We have used Plummer models since they are easy to work with and it is also possible to validate the outcome of the simulations using analytical methods. We do not expect that other initial density distributions such as \citet{King} models will affect the final results significantly, as the exact details of any initial density distribution will be quickly wiped out by violent relaxation as a result of significant mass-loss in the simulated star clusters.

\par The central gas cloud and SG stars have the same degree of concentration with respect to FG stars in our simulations, i.e. $a_2/a_1=a_g/a_1 \in \{0.1, 0.2\}$ where $a_1$, $a_2$ and $a_g$ are the Plummer scale radii of FG stars, SG stars and the gas cloud respectively. Table \ref{tab:initialConditions} summarizes the initial conditions of our simulations.

\par The time-scale of our simulations is short compared to the clusters relaxation times so the effect of stellar evolution, mass segregation, etc. can be neglected. As a result all particles in our simulations have equal masses which are constant throughout the whole simulation. All time-scales in our simulations are expressed in terms of the the initial crossing time of the cluster which is defined to be:
\begin{equation}
	T_{\rm cr} \equiv \frac{2r_h}{\sigma_v}
\end{equation}
Where $r_h$ is the initial half-mass radius of all stars (which is approximately equal to 1.20 times the Plummer radius of FG stars for the values adopted in Table \ref{tab:initialConditions}, i.e. $r_h\approx1.20a_1$) and $\sigma_v$ is the initial velocity dispersion of stars calculated from the virial theorem in the presence of gas.

\par We start with clusters which are initially in virial equilibrium and then remove the gas according to the following equation
\begin{equation}
	M_g (t)=\begin{cases} M_g (0)\exp\left(-\displaystyle\frac{t-t_0}{\tau}\right) &  t>t_0 \\ M_g (0) & t\leq t_0 \end{cases}
\end{equation}
where $t$ is the simulation time, $t_0$ is the amount of time that we wait before removing the gas\footnote{The reason that we don't remove gas at $t=0$ is that we want to measure the dynamical properties of the simulated clusters in the first few crossing times when the cluster is still in equilibrium.} and is set to be equal to $5$ crossing times in all simulations and $\tau=T_\Exp/T_{\rm cr}$ is the ratio of gas expulsion time-scale to the initial crossing time. We change $\tau$ in the range $10^{-2}$ to $10^4$ on a logarithmic scale, corresponding to instantaneous and adiabatic gas expulsion respectively.

\par The initial mass of gas $M_{\rm gas}(0)$ is parameterized by a parameter $\eta$ which is the ratio of the initial mass of gas divided by the initial mass of FG stars, i.e.
\begin{equation}
	\eta=\frac{M_g(0)}{M_1(0)}
\end{equation}
where $\eta$ varies from 0 to 2.00 in steps of 0.02 in our simulations. 

\par All simulated clusters are in a Galactic tidal field which is modelled using the near-field approximation \citep{Aarseth} and implemented by writing the equations of motions of stars in a right-handed rotating coordinate system whose origin is initially centered on the cluster and $x$ and $y$ axes point toward the Galactic anti-center and the direction of orbital motion of the star cluster respectively, assuming that the star cluster is moving in the $x-y$ plane. The equation of motion for a star in such a coordinate system is given by
\begin{equation}\label{eq:totalAcceleration}
	{\ddot{\mathbf r}}_i ({\rm total})={\ddot{\mathbf r}}_i ({\rm stars})+{\ddot{\mathbf r}}_i ({\rm gas})-2{\mathbf \Omega}\times{\dot{\mathbf r_i}}+\Omega^2(3x_i{\mathbf e_x}-z_i{\mathbf e_z})
\end{equation}
Where ${\ddot{\mathbf r}}_i ({\rm stars})+{\ddot{\mathbf r}}_i ({\rm gas})$ is the acceleration of each star due to the total gravitational force of other stars and the gas cloud which are calculated using the following equations
\begin{equation}
	{\ddot{\mathbf r}}_i ({\rm stars}) = \sum_{j=1, j\neq i}^{j=N}\frac{Gm_j}{\bigg(|{\mathbf r_j}-{\mathbf r_i}|^2+\epsilon^2\bigg)^{3/2}}({\mathbf r_j}-{\mathbf r_i})
\end{equation}

\begin{equation}
	{\ddot{\mathbf r}}_i ({\rm gas}) = -\frac{GM_g(t)}{\bigg(r_i^2+a_g^2\bigg)^{3/2}}{\mathbf r_i}
\end{equation}
where $\epsilon$ is the softening parameter that we have introduced in our simulations and it is equal to the minimum distance between stars in the central region of the cluster.
The third and forth terms on the right-hand side of Eq. \eqref{eq:totalAcceleration} are the Coriolis and centrifugal force combined with the tidal forces respectively and ${\mathbf \Omega}=\Omega{\mathbf e_z}$ is the angular velocity of the cluster around the Galactic center.

\par In our simulations the strength of tidal field is parameterized by the ratio of the tidal radius of each cluster to the Plummer radius of FG stars $r_t/a_1$. We vary $r_t/a_1$ from 5 to 40 in steps of 5, where a large value of $r_t/a_1$ means a weak tidal field. We also did one set of simulations for $r_t/a_1=\infty$ ($\Omega=0$) which corresponds to isolated clusters. $r_t$ is related to the total cluster mass $M_\star(t)+M_{\rm gas}(t)$ and $\Omega$ via the following equation \citep{Giersz}:
\begin{equation}\label{eq:rt}
	r_t(t)=\left(G\frac{M_\star(t)+M_g(t)}{3\Omega^2}\right)^{1/3}
\end{equation} 

\par As a result, all clusters in our grid can be modelled by only 4 parameters $\eta$, $r_t/a_1$, $\tau$ and $\lambda=a_2/a_1$ (see Table \ref{tab:initialConditions}). Our simulations can be thought of as a generalized version of the \citet{Baumgardt07} models who considered the effect of gas expulsion on a single stellar population. 

\par All simulations in our grid are run for 555 initial crossing times which was found to be enough for clusters with $\tau<10^3$ to end up in a quasi-equilibrium state after which we determined the mass, half-mass radius and number ratio of SG stars. For $\tau>10^3$ the gas expulsion is adiabatic which only affects the cluster in long term and is dealt with in Section \ref{sec:observations}.

\par We ran all the simulations on the Green II GPU supercomputer at Swinburne University of Technology.


\begin{figure*}
	\includegraphics[width=0.85\textwidth]{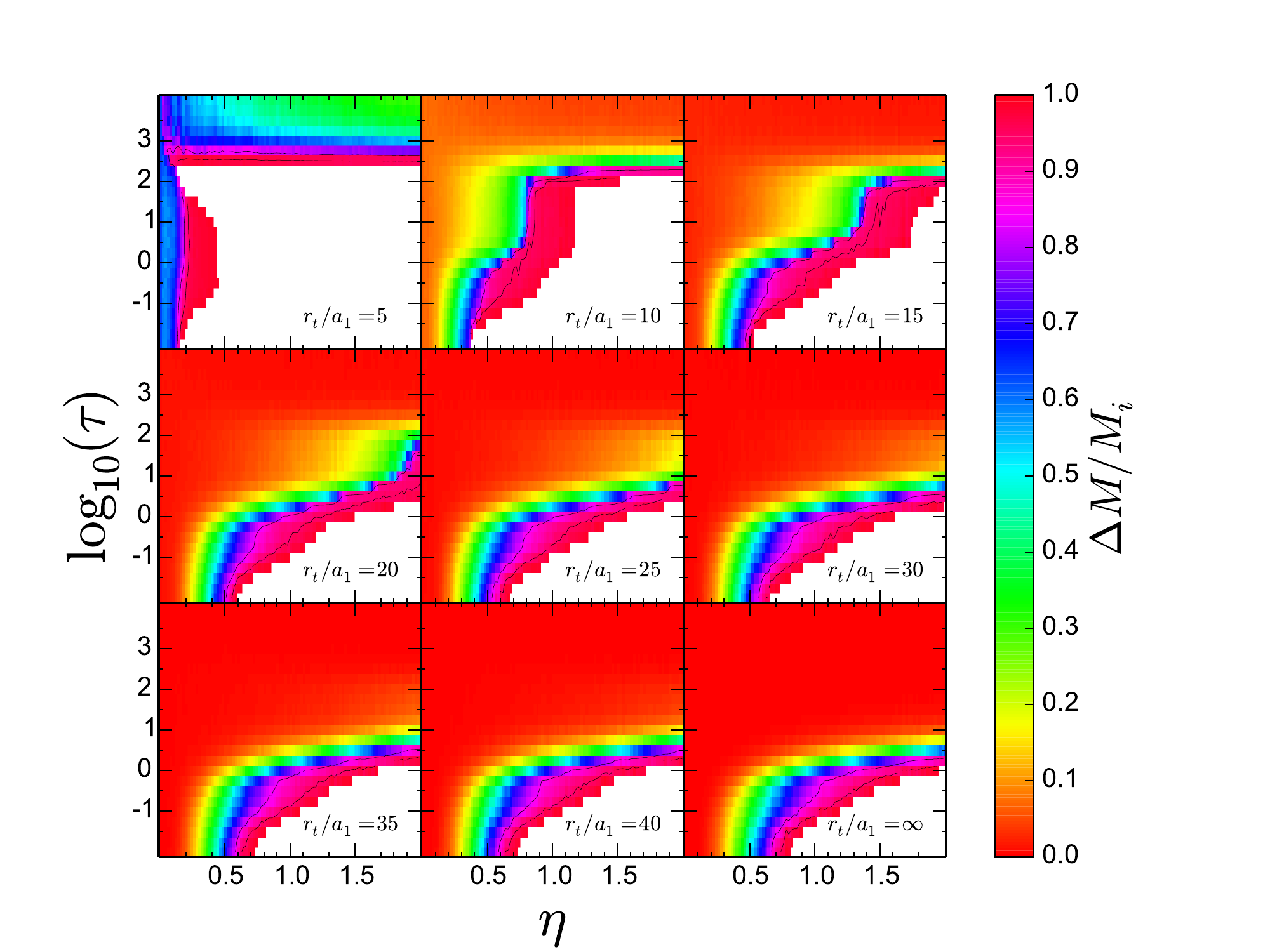}
	\includegraphics[width=0.85\textwidth]{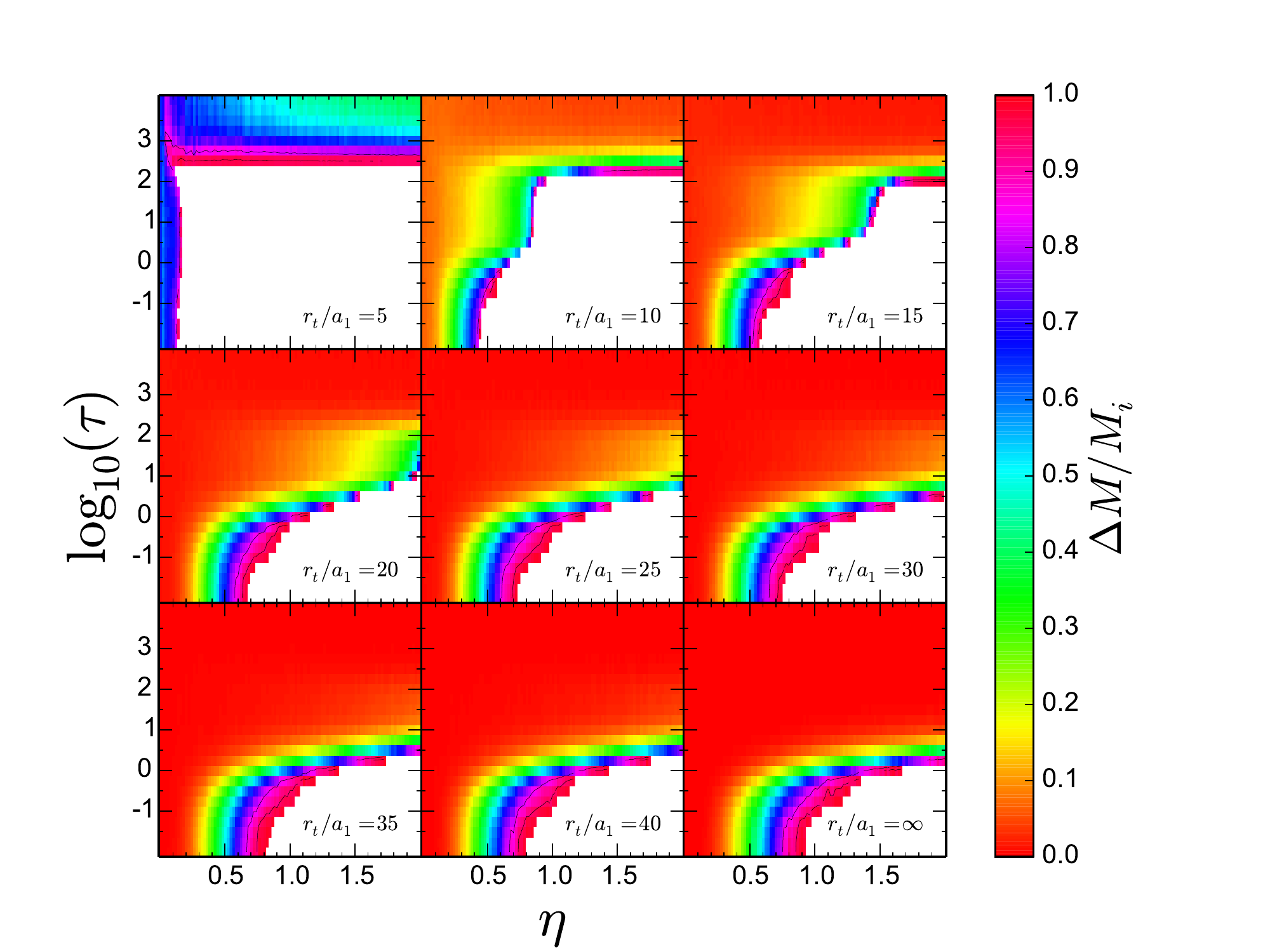}  
	\caption{Mass-loss ${\Delta M}/{M_i}$ as a function of gas fraction $\eta$ and gas expulsion time-scale $\tau$ for different strengths of the tidal field (${r_t}/{a_1}=5, 10, 15, 20, 25, 30, 35, 40, \infty$) and concentration of SG stars $\lambda$. The top and the bottom panels correspond to $\lambda=0.1$ and $\lambda=0.2$ respectively. The region enclosed by solid black lines corresponds to $90\pm5\%$ mass-loss. The white filled area in the lower right corner of each plot shows the region where all clusters are totally destroyed.}
	\label{fig:grid1}
\end{figure*}

\begin{figure*}	
	\includegraphics[width=0.85\textwidth]{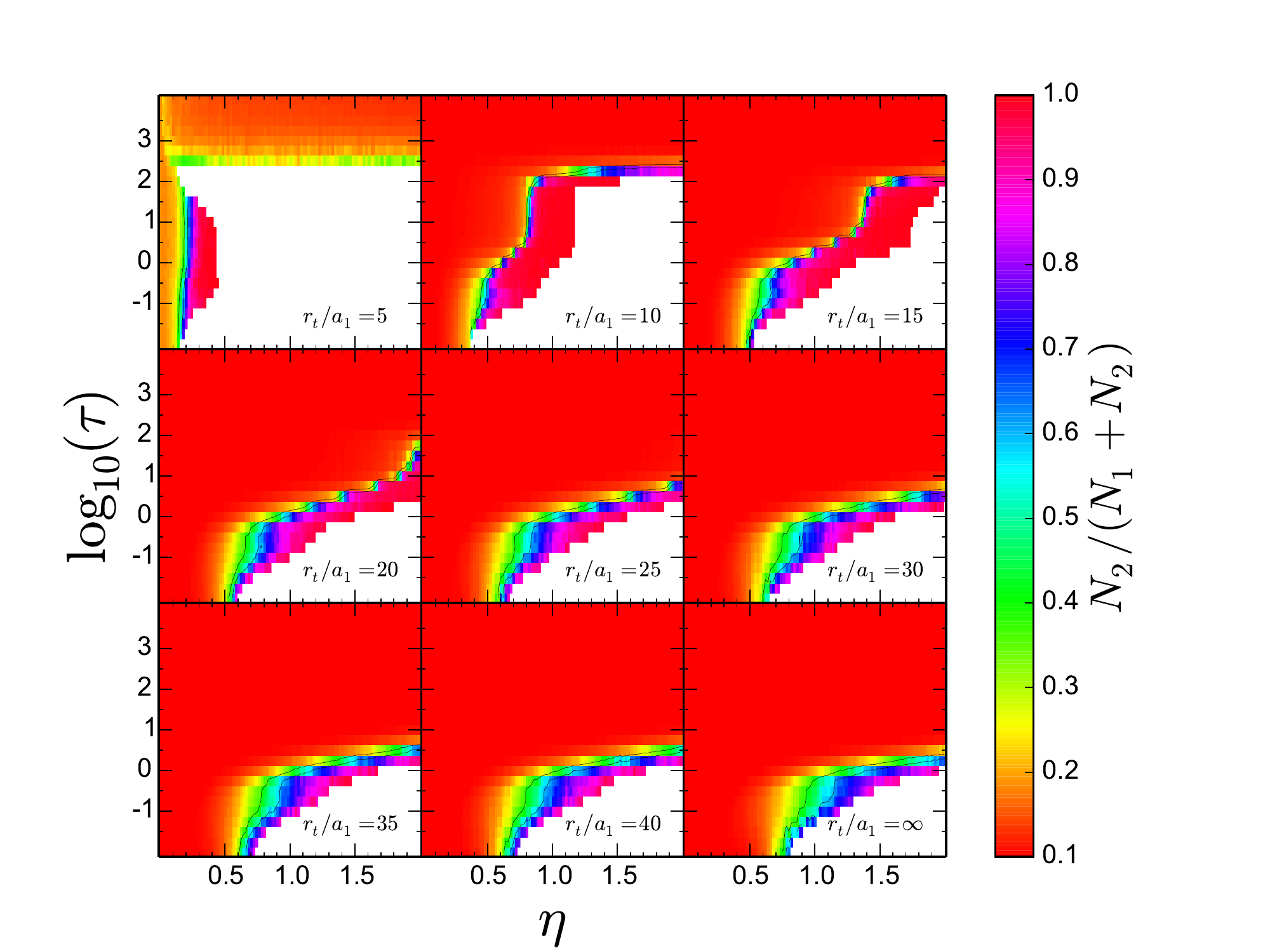}
	\includegraphics[width=0.85\textwidth]{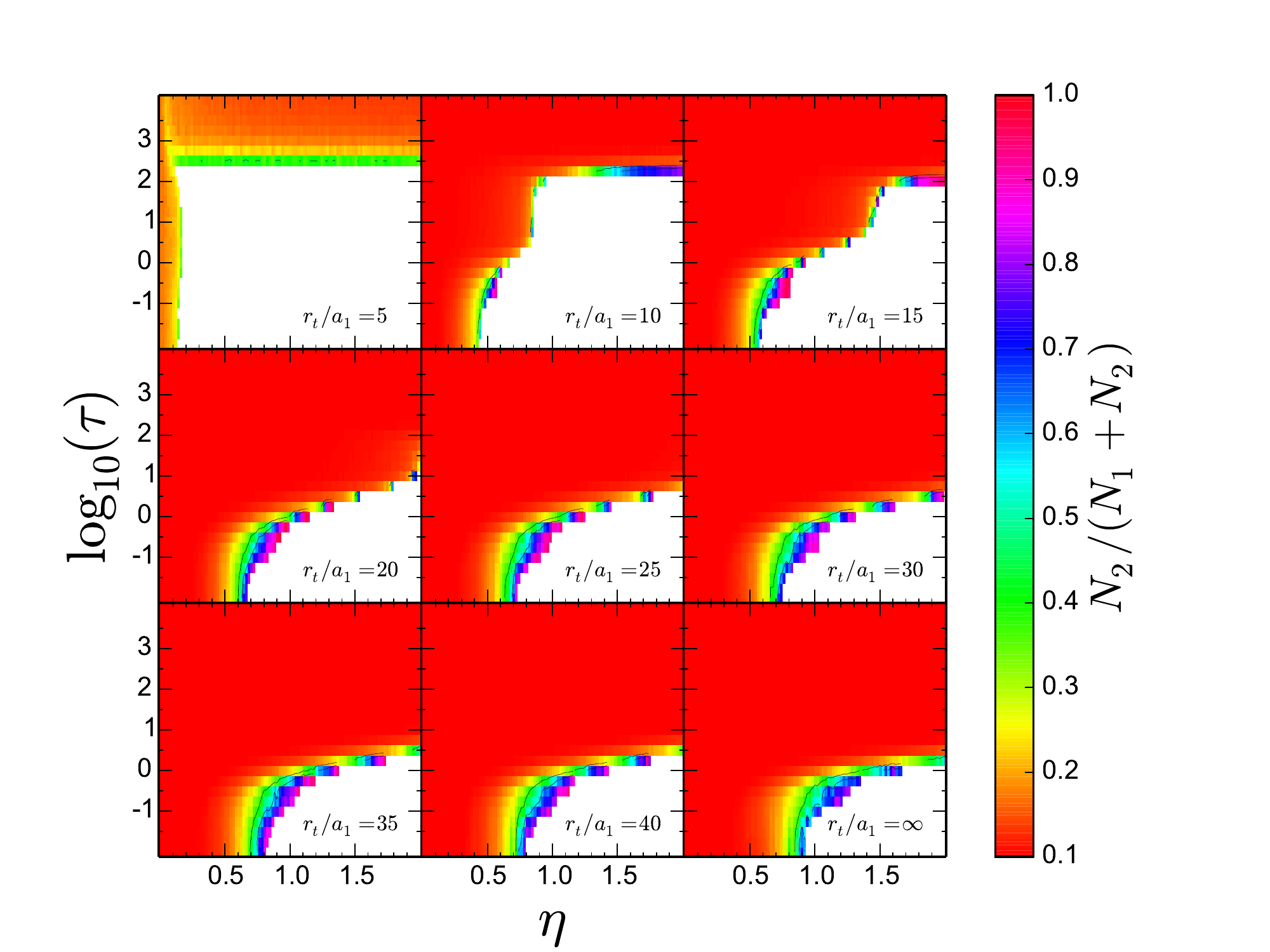}
	\caption{Same as Fig. \ref{fig:grid1} but for the number ratio of SG stars ${N_2}/({N_1+N_2})$. The region enclosed by solid black lines corresponds to a number ratio of $50\pm10\%$ for SG stars.}
	\label{fig:grid2}
\end{figure*}

\begin{figure*}	
	\includegraphics[width=0.85\textwidth]{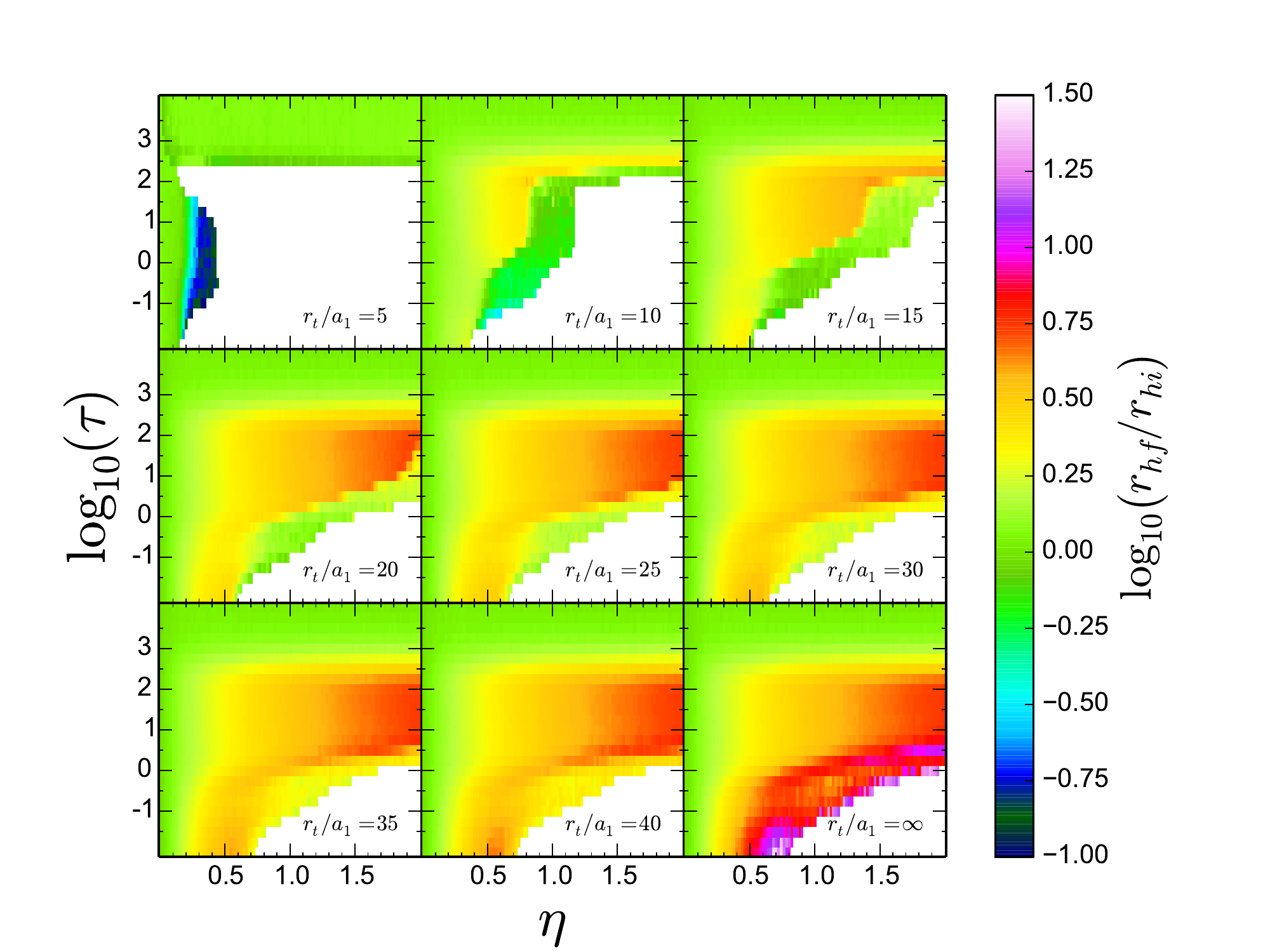}
	\includegraphics[width=0.85\textwidth]{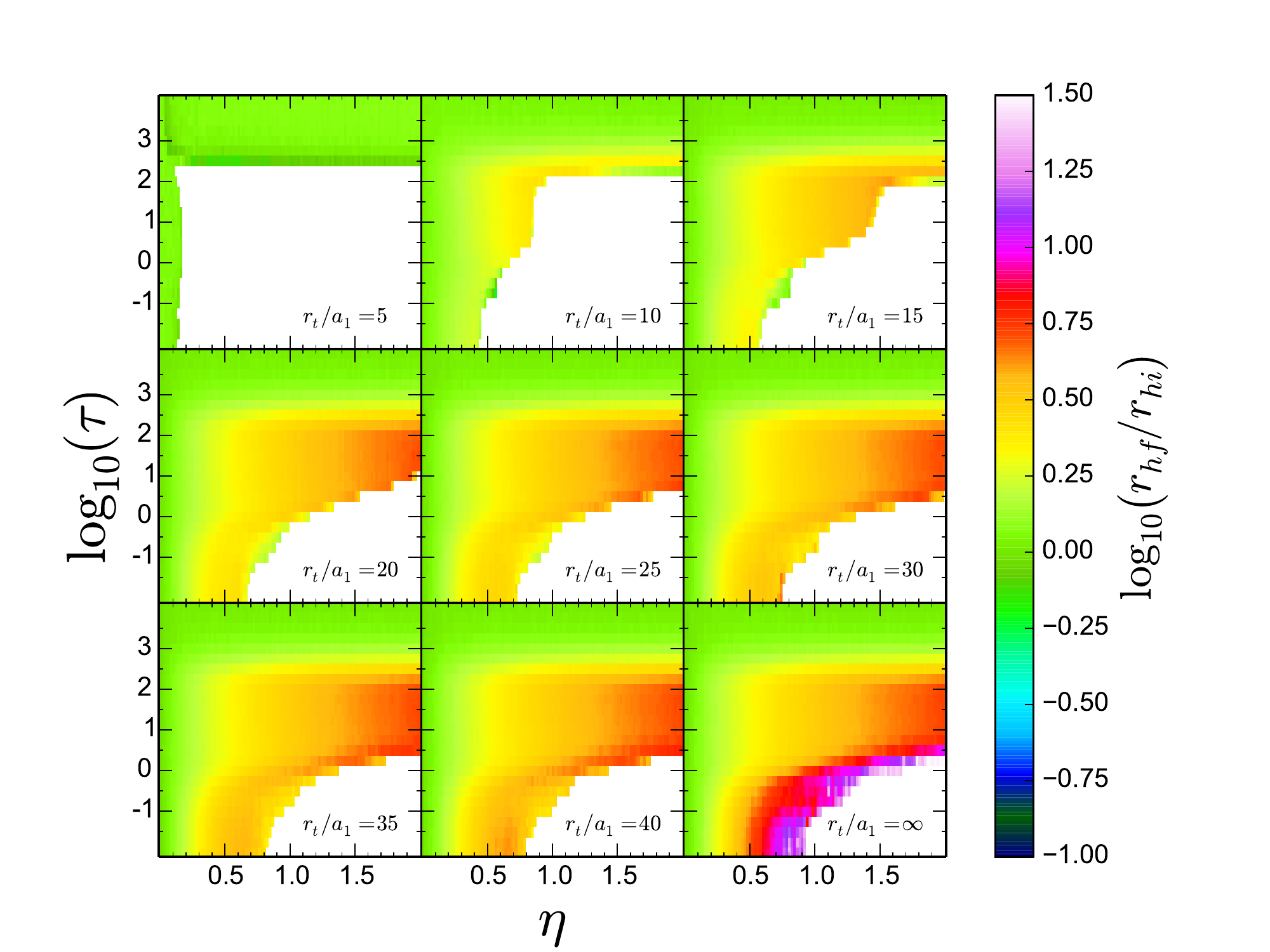}
	\caption{Same as Fig. \ref{fig:grid1} but for the logarithm of the expansion factor of the cluster $\log_{10}\left({r_{hf}}/{r_{hi}}\right)$, for $\lambda=0.1$ (top) and $\lambda=0.2$ (bottom).}
	\label{fig:grid3}
\end{figure*}

\section{Results}\label{sec:results}
We record the properties of our model clusters throughout the simulation. In particular, we find unbound stars using an iterative algorithm as described below:
\begin{enumerate}
	\item Find the coordinate of the cluster density center using the \citet{vonHoerner63} method and the unbiased density estimator of \citet{Casertano} for the 10th nearest neighbours of each star, i.e. $j=10$ in equation II.2 of \citet{Casertano}.
	\vspace{.2cm}
	\item Using Eq. \eqref{eq:rt}, calculate the instantaneous tidal radius of the cluster $r_t(t)$ as a function of the remaining cluster mass.
	\vspace{.2cm}
	\item Find the stars whose distances are larger than the tidal radius calculated in the previous step and mark them as unbound stars. For isolated clusters ($r_t/a_1=\infty$), use the total energy of each star as the selection criterion.
	\vspace{.2cm}
	\item Subtract the mass of unbound stars from the total mass of the cluster.
	\vspace{.2cm}
	\item Repeat the previous steps until all bound stars reside within the tidal radius or all stars are designated as unbound stars (i.e. total disruption).
\end{enumerate}

\par Using the above algorithm we calculate the mass-loss, number ratio of SG stars and expansion factor for the model clusters at each instant of the simulation. The outcome of our simulations is shown in Fig. \ref{fig:grid1} to \ref{fig:grid3} which depict the mass-loss ${\Delta M}/{M_i}$, number ratio of SG stars ${N_2}/({N_1+N_2})$ and logarithm of expansion factors $\log_{10}\left({r_{hf}}/{r_{hi}}\right)$ as a function of gas fraction $\eta$ and the ratio of the gas expulsion time-scale to the initial crossing time $\tau$ for different tidal radii ratios (${r_t}/{a_1}=5, 10, 15, 20, 25, 30, 35, 40, \infty$) and $\lambda=(0.1, 0.2)$. In these figures, the region enclosed in solid lines corresponds to $90\pm5\%$ mass-loss and a value of $50\pm10\%$ for the fraction of SG stars. The white filled area in the lower right corner of each plot is the total disruption zone in which all clusters will be totally destroyed as a result of significant mass-loss. The region between the black solid lines represents the set of initial conditions which match the observations. One can see that the width of this region increases in stronger tidal fields and for higher concentrations of SG stars (e.g. $\lambda=0.1$).

\par The trend that we see in these figures can be explained as follows: First, the loss of gravitational potential is greater for clusters with a higher gas fraction. Second, very short gas expulsion times-scales do not allow loosely-bound stars (mainly FG stars) to compensate for the loss of gravitational potential energy and go into an equilibrium so they leave the cluster after a few crossing times, whereas in models with longer gas expulsion time-scales stars have enough time to gradually expand their orbits and remain in a quasi-equilibrium state without crossing the tidal radius and escaping from the cluster. Third, FG stars have a lower concentration than SG stars and when the gas expulsion is instantaneous, the cluster preferentially loses more FG stars, while in the adiabatic case, many of FG stars will be retained in the cluster. Fourth, clusters that lose a substantial number of stars have smaller tidal radii and have shrunk in size, hence their expansion factor is less than 1.0 and decreases with mass-loss. As a result, the mass-loss is expected to be much more extreme for clusters with higher gas fractions and shorter time-scales and such clusters must show a higher number ratio of SG stars and relatively lower expansion factors. In addition, there should be a region in the parameter space in which all clusters will be totally disrupted. The outcome of our \Nbody simulations are consistent with \citet{Decressin10} who analysed the \Nbody models of \citet{Baumgardt07}. 

\par As it is inferred from Fig. \ref{fig:grid1} to \ref{fig:grid3}, the initial conditions which meet the observational criteria occupy a very narrow strip in the parameter space. In addition, this region is very close to the total disruption zone, meaning that if we slightly change the initial conditions, we will either end up in the total disruption zone (SG fraction $\sim 100\%$) or the region which is far from the observed clusters (SG fraction $\sim 10\%$). We did a MC analysis as explained in Section \ref{sec:observations} in order to find the physical initial conditions of GCs in terms of cluster mass, half-mass radius and gas expulsion time-scale.


\begin{table}
	\centering
	\caption{Range of the initial parameters used in the MC simulations.}
	\label{tab:MCParameters}
	\begin{minipage}{85mm}	
		\begin{tabular}{ccc}
			\hline
			\hline
			Parameter & Range & Steps \\
			\hline
			$\overline{\log \left(\frac{M_\star}{\Msun}\right)}$ & [5.4, 6.5] & 0.05 \\ \\
			$\overline{\eta}$ & [0.7, 1.5] & 0.05 \\ \\
			$\overline{\log \left(\frac{T_\Exp}{yr}\right)}$ & [3.2, 5.2] & 0.05 \\ \\
			$\overline{\log \left(\frac{r_{h}}{\pc}\right)}${\footnote{This parameter is only relevant in models with mass-independent radii.}} & [-0.25, 0.5] & 0.05 \\   \\
			$\sigma_{\log \left({T_\Exp}/{\yr}\right)}$ = $\sigma_{\log (M_\star/\Msun)}$ = $\sigeta$ & [0.25, 0.75] & 0.25 \\ \\
			$\sigma_{\log \left({r_{h}}/{\pc}\right)}$ & [0.15, 0.45] & 0.15 \\    
			\hline 
		\end{tabular}
	\end{minipage}
\end{table}

\begin{table*}
	\centering 	
	\caption{Outcome of the MC simulations. We have simulated 1000 clusters for each set of the initial parameters. Each row shows the best-fitting model in terms of the $D$ parameter which is a measure of the goodness of fit and defined by Eq. \eqref{eq:D}. All models in this table have sample mean values of $50\pm5\%$ for the fraction of SG stars. $r$ refers to the value of the anti-correlation between the cluster mass and the fraction of SG stars. The values reported in this table are the means of 20 samples with different random seed numbers. The best-fitting models show statistical fluctuations within $\pm0.1$ for $\overline{\log \left({M_\star}/{\Msun}\right)}$,  $\overline{\eta}$ and $\overline{\log \left({T_\Exp}/{\yr}\right)}$ which can be inferred as an error-bar on the best-fitting parameters. }
	\label{tab:MCSimulations}
	\begin{tabular}{ccccccccc}
		\hline
		\hline
 		Tidal Field & $\lambda$ & $\overline{\log \left(\frac{M_\star}{\Msun}\right)}\pm\sigma$ & $\overline{\eta}\pm\sigma$ & 
	     $\overline{\log \left(\frac{T_\Exp}{\yr}\right)}\pm\sigma$ & $\overline{\log \left(\frac{r_{h}}{\pc}\right)}\pm\sigma$ & $r$ & $D \left(\times 10^{-2}\right)$  \\
	     \hline
		{\bf mass-dependent radii} & & & & & & & &\\
		{\small MD-RG2.0} & 0.1 & $5.95\pm0.75$ & $1.25\pm0.50$ & $4.45\pm0.50$ & -- & -0.50 & 1.34 \\ 
		{\small MD-RG2.0} & 0.2 & $6.05\pm0.50$ & $1.30\pm0.25$ & $3.95\pm0.75$ & -- & -0.72 & 1.86 \\ 
		{\small MD-RG4.0} & 0.1 & $5.60\pm0.50$ & $1.30\pm0.50$ & $4.30\pm0.50$ & -- & -0.70 & 1.33 \\ 
		{\small MD-RG4.0} & 0.2 & $5.65\pm0.50$ & $1.15\pm0.25$ & $3.60\pm0.75$ & -- & -0.72 & 1.61 \\ 
		{\small MD-RG8.5} & 0.1 & $5.65\pm0.50$ & $1.10\pm0.50$ & $4.20\pm0.25$ & -- & -0.72 & 1.34 \\ 
		{\small MD-RG8.5} & 0.2 & $5.60\pm0.50$ & $1.20\pm0.25$ & $4.15\pm0.50$ & -- & -0.70 & 1.58 \\ \\
		{\bf mass-independent radii} & & & & & & & &\\
		{\small MI-RG2.0} & 0.1 & $6.10\pm0.25$ & $0.95\pm0.25$ & $3.90\pm0.25$ & $0.10\pm0.30$ & -0.85 & 0.87 \\ 
		{\small MI-RG2.0} & 0.2 & $6.15\pm0.25$ & $1.00\pm0.25$ & $3.40\pm0.25$ & $0.05\pm0.30$ & -0.86 & 0.99 \\ 
		{\small MI-RG4.0} & 0.1 & $6.15\pm0.25$ & $0.95\pm0.25$ & $3.55\pm0.25$ & $0.10\pm0.30$ & -0.83 & 0.90 \\ 
		{\small MI-RG4.0} & 0.2 & $6.20\pm0.25$ & $1.05\pm0.25$ & $3.40\pm0.25$ & $-0.05\pm0.30$ & -0.87 & 0.99 \\ 
		{\small MI-RG8.5} & 0.1 & $6.05\pm0.25$ & $0.95\pm0.25$ & $3.65\pm0.25$ & $0.05\pm0.30$ & -0.84 & 0.87 \\ 
		{\small MI-RG8.5} & 0.2 & $6.10\pm0.25$ & $1.05\pm0.25$ & $3.40\pm0.25$ & $-0.05\pm0.30$ & -0.86 & 0.97 \\ 		
		\hline
		\hline 
	\end{tabular}	
\end{table*}

\section{Monte Carlo Analysis and Comparison With Observations}\label{sec:observations}
In this section we describe the details of our MC analysis on the initial conditions of star clusters. We make different sets of initial conditions for star clusters and we feed these initial conditions into our grid to find the final conditions and compare them with observations of Galactic GCs. We change the initial distribution until the best match with observations is found. We have performed our MC analysis for $\lambda=0.1$ and $0.2$ separately. We adopt a log-normal distribution for the initial distribution of the cluster stellar mass \citep{Parmentier07, Parmentier08} parameterized by a mean value and standard deviation of $\overline{\log \left({M_\star}/{\Msun}\right)}$ and $\sigma_{\log (M_\star/\Msun)}$ respectively. We assume similar normal and log-normal distributions for the gas fraction and gas expulsion time-scale with mean values of $\overline{\eta}$ and $\overline{\log \left({T_\Exp}/{\yr}\right)}$ and standard deviations equal to $\sigeta$ and $\sigma_{\log \left({T_\Exp}/{\yr}\right)}$. We assume that the initial half-mass radii and the initial masses of GCs are related via the following initial mass-radius relation derived by \citet{Gieles10}
\begin{equation}\label{eq:massRadius}
	\log\left(\frac{r_{h}}{\pc}\right)=-3.5650+0.615\log\left(\frac{M}{\Msun}\right)
\end{equation}
For comparison, we have also done one set of MC simulations by relaxing the mass-dependent constraint on radii and replacing it with a log-normal distribution to see how it affects the final results.

\par Tidal radii of GCs depend on the environment in which they form which is unknown. Possible choices are (1) GCs have formed in an environment similar to the present-day Milky Way, when most of its mass was already in place or (2) they formed in satellite galaxies of the Milky Way with many of them being disrupted and merged with the Milky Way \citep{Prieto} and some survived like the LMC and Fornax dwarf galaxy.
For the first case we assume that our clusters are in a Galactic field with a constant circular velocity of $\VG=220\kms$ at a distance $\RG$ from the Galactic center. The tidal radius can then be determined using the following equation (Eq. 1 of \citealt{Baumgardt03})
\begin{equation}\label{eq:rt}
	r_t=\left(\frac{GM}{2\VG^2}\right)^{1/3}\RG^{2/3}
\end{equation}
\par In our MC analysis we have considered three cases of $\RG=2$, 4 and $8.5\kpc$ (solar neighbourhood) corresponding to strong, moderate and weak tidal fields in the present-day Milky Way. We will refer to these cases as MD-RG2.0, MD-RG4.0 and MD-RG8.5 for mass-dependent radii, as given by Eq. \eqref{eq:massRadius}, and MI-RG2.0, MI-RG4.0 and MI-RG8.5 for mass-independent radii.

\par If GCs formed in dwarf galaxies, they would have tidal radii which are comparable to the $\RG=8.5\kpc$ case. One can take LMC with $\RG=4\kpc$, $\VG=70\kms$ \citep{Alves} or Fornax with $\RG=0.5\kpc$, $\VG=10\kms$ \citep{Strigari} as an example, where $\RG$ refers to the radius of the galaxy and $\VG$ is the circular velocity at $\RG$. Using Eq. \eqref{eq:rt}, the tidal radius of a cluster with a mass of $M=10^6\Msun$ in such galaxies is about $190$ and $175\pc$ respectively, close to the value of $150\pc$ for $\RG=8.5\kpc, \VG=220\kms$.
As a result our three choices of the tidal strength are sufficient to represent different environments in which GCs might have formed.

\par Given the tidal radius and the initial half-mass radius of each cluster, the ratio of $r_t/r_h$ and consequently $r_t/a_1$ can be calculated. As a result all the required initial conditions $(\eta, \tau, r_t/a_1)$ to identify our model clusters in the \Nbody grid will be uniquely determined and we can find the final properties of the clusters by interpolation between the values of the grid. In order to interpolate in our grid we need to assume that mass-loss, fraction of SG stars and expansion factor in the total disruption zone (white area in Fig. \ref{fig:grid1} to \ref{fig:grid3}) are equal to 1.0, 1.0 and 0.0 respectively. These assumptions are based on the fact that the clusters with a mass-loss of about $99\%$ are mainly composed of SG stars and have expansion factors less than 1.0, as explained in Section \ref{sec:nbodySimulations}. We would like to stress that we interpolate in a 3D parameter space $(\eta, \tau, r_t/a_1)$, so although all clusters are in the same tidal field they don't have the same $r_t/a_1$.

\par In our analysis, we also consider the late-time adiabatic expansion of clusters as a result of the remnant gas expulsion (for clusters with $\tau>10^3$) and also stellar evolution induced mass-loss by scaling the radii of all clusters according to the mass-radius relation of \citet{Hills} which states that the radius of a cluster inversely scales with its mass, i.e. $r_{h}\propto M^{-1}$, assuming that the cluster remains in virial equilibrium after the initial significant mass-loss. We have used the AMUSE\footnote{AMUSE (Astrophysical Multipurpose Software Environment) is available at http://amusecode.org} code \citep{PortegiesZwart09, Pelupessy, PortegiesZwart13} and analytic stellar evolution models from \citet{Hurley} to find the stellar evolution induced mass-loss for each cluster, which is on average equal to $\sim 30\%$ of the initial mass of the cluster, calculated for interval $t_{\rm end} < t < 13.8\,{\rm Gyr}$, where $t_{\rm end}$ is the age of the cluster in physical units at the end of \Nbody simulation. The IMF of FG and SG stars is a \citet{Kroupa} IMF which extends to $100\Msun$ and $8\Msun$ respectively. SG stars cannot be more massive than $8\Msun$, otherwise they will explode as SN and in the AGB scenario this will change the iron abundance of SG stars which is not consistent with observations \citep{DErcole08}.

\begin{figure*}  		
	\includegraphics[width=0.33\textwidth]{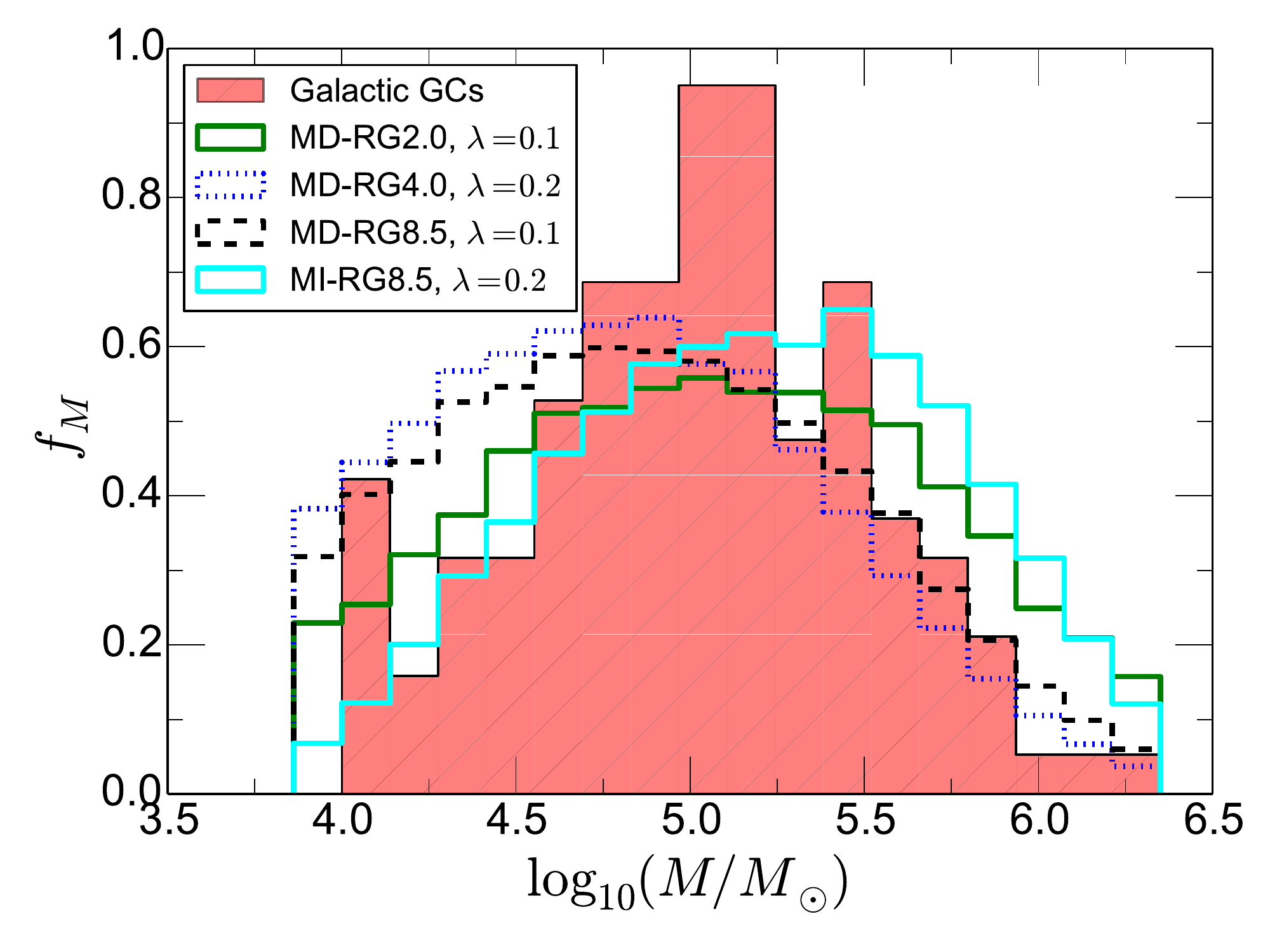}
	\includegraphics[width=0.33\textwidth]{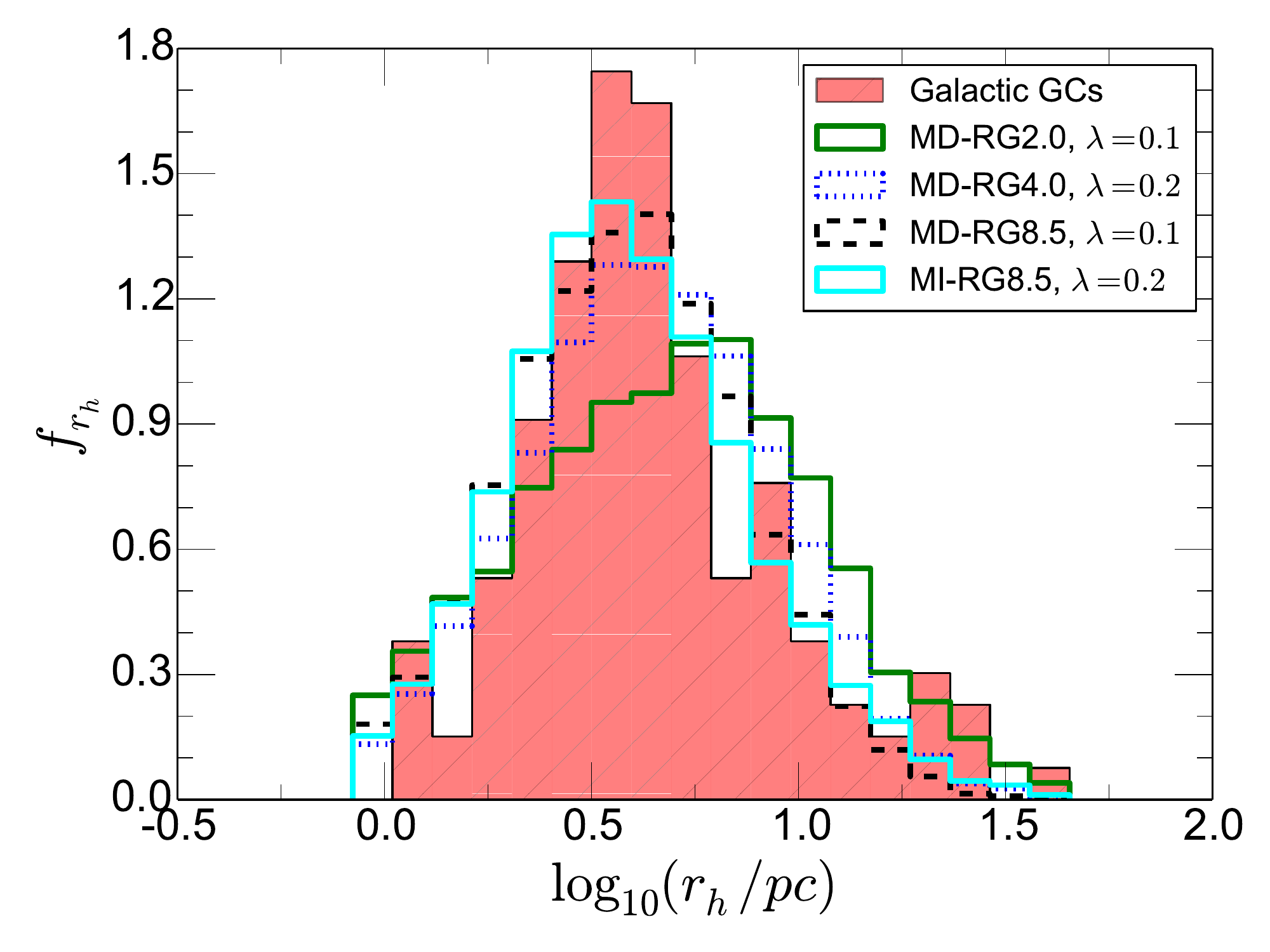}
	\includegraphics[width=0.33\textwidth]{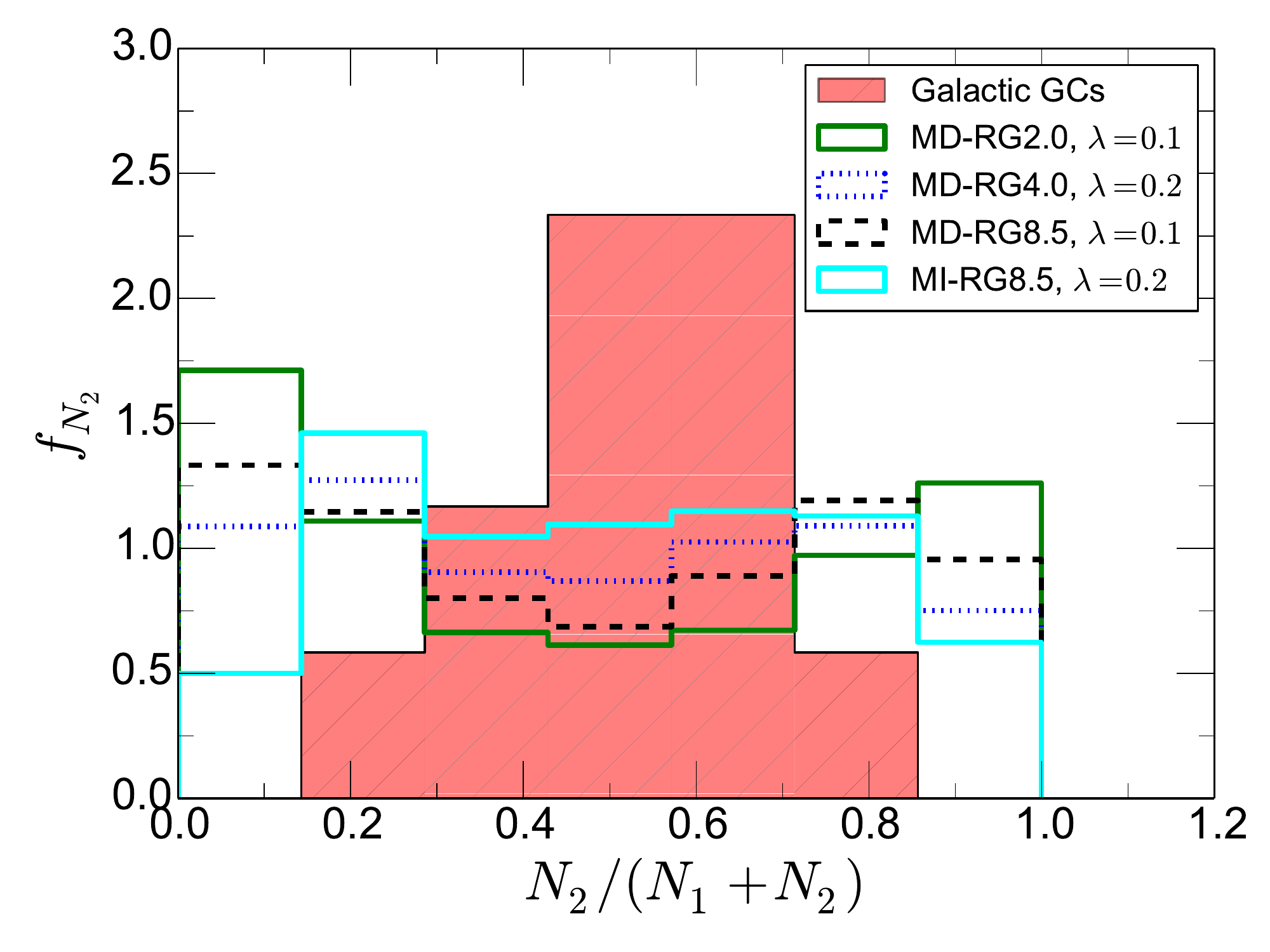}		
	\caption{Comparison of the distribution of cluster masses (left), half-mass radii (middle) and the fraction of SG stars (right) for observed clusters (hatch-filled) and four of our best-fitting models listed in Table \ref{tab:MCSimulations}. Cluster masses and radii are taken from the most recent version of \citet{Harris} and fraction of SG stars are taken from \citet{DAntona08}.}
	\label{fig:distributions}
\end{figure*}

\begin{figure*}  		
	\includegraphics[width=0.33\textwidth]{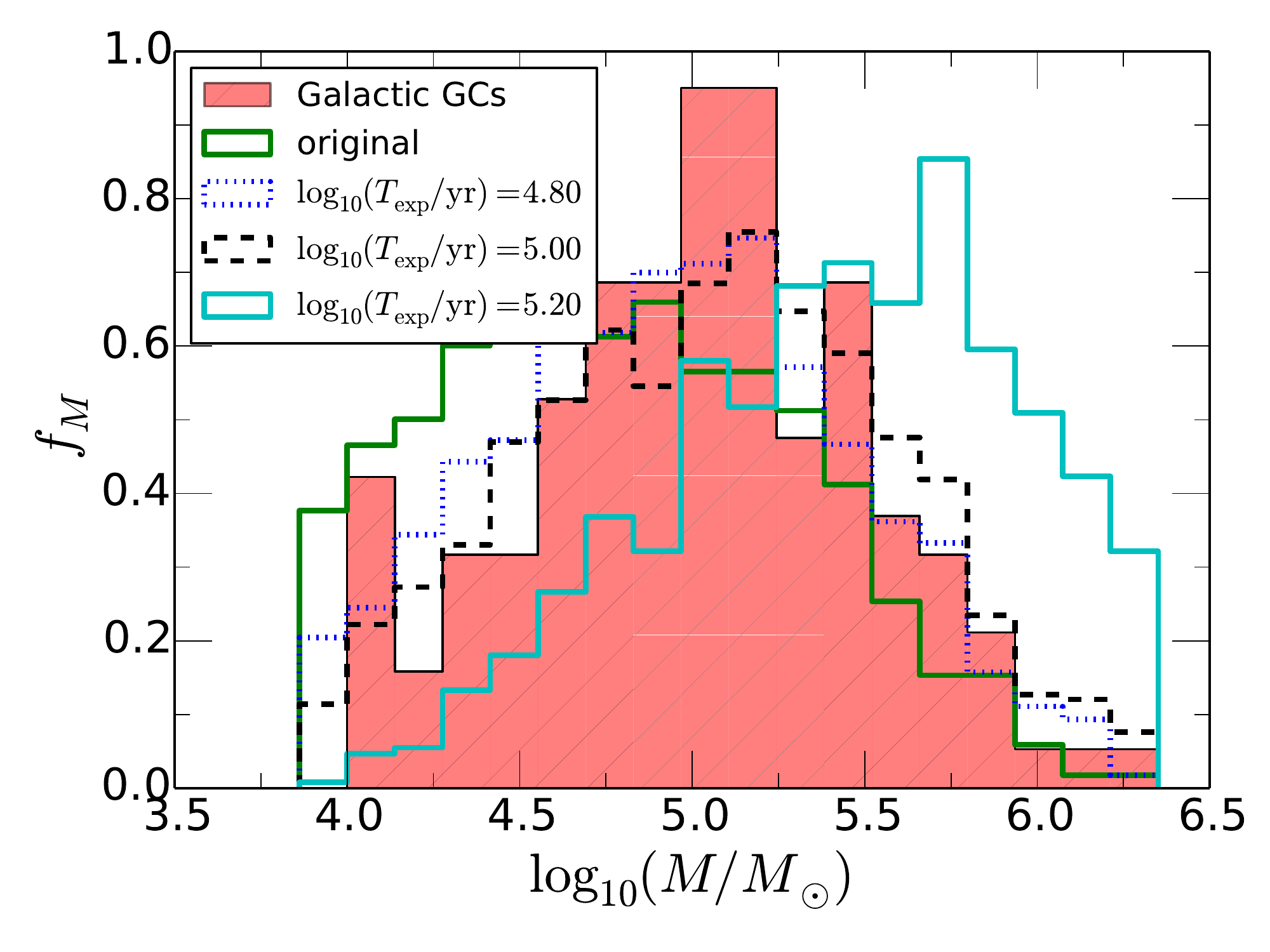}
	\includegraphics[width=0.33\textwidth]{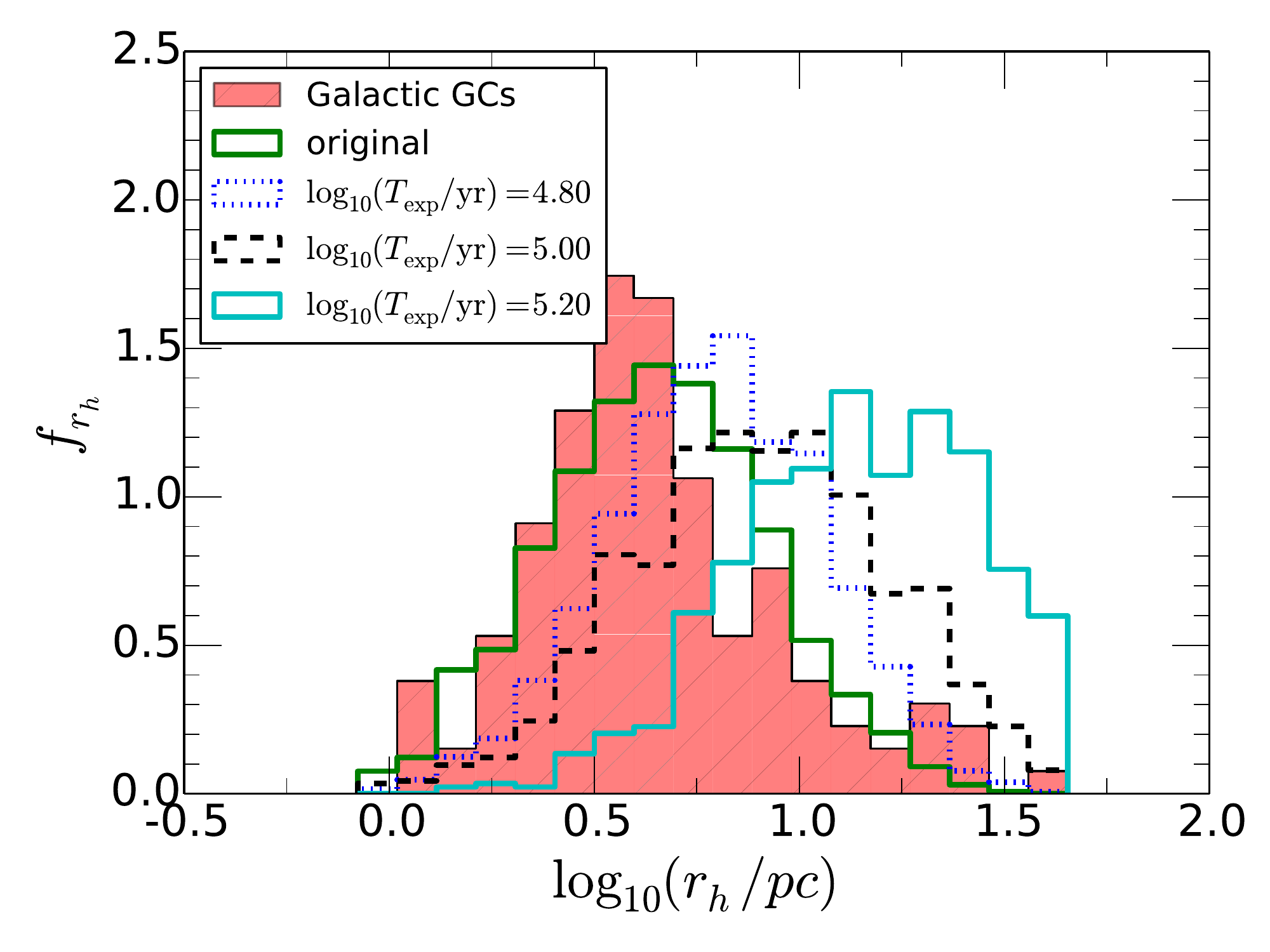}
	\includegraphics[width=0.33\textwidth]{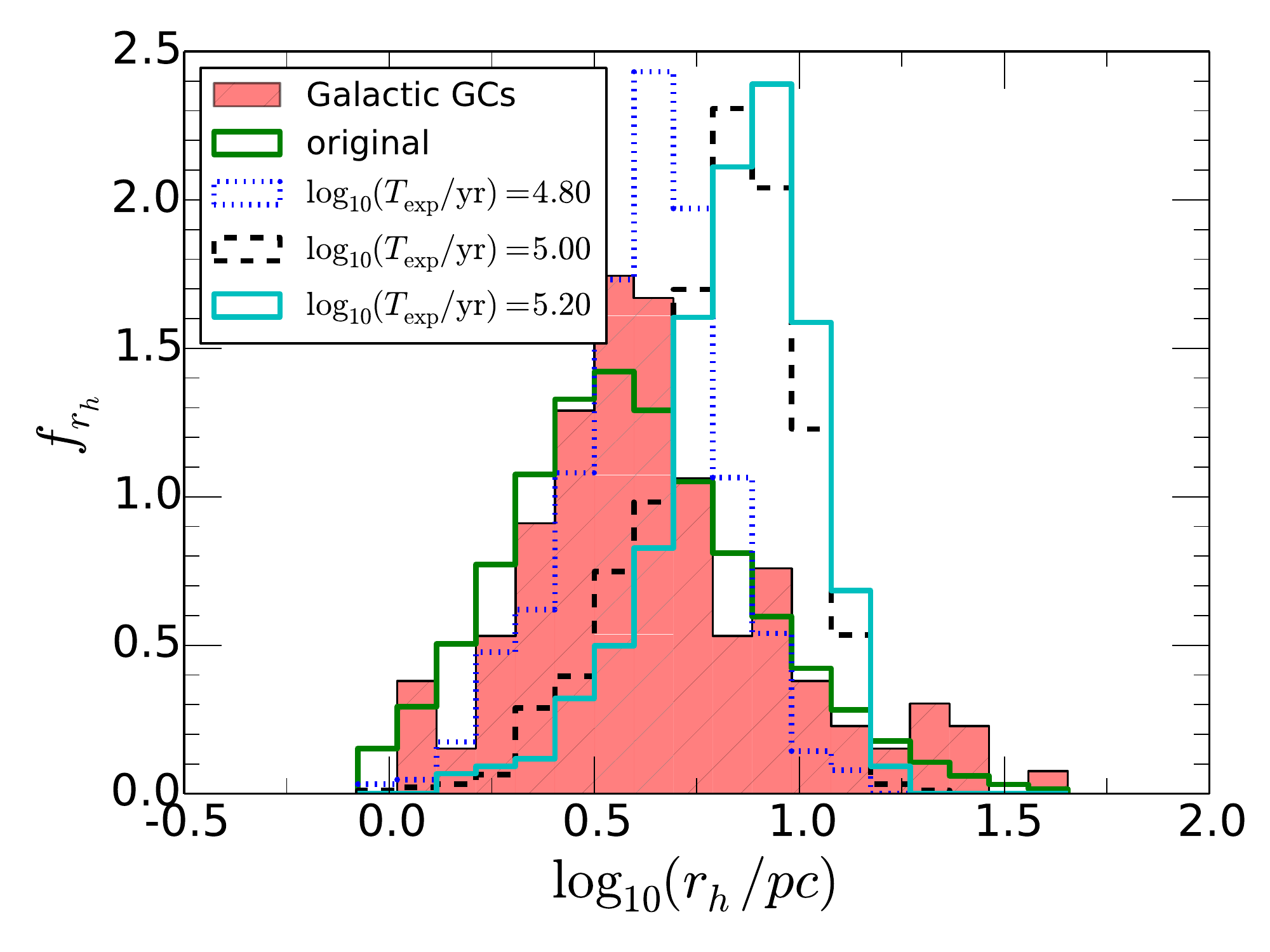}			
	\caption{Distribution of mass and half-mass radii of MD-RG8.5 (left and middle) and MI-RG8.5 (right) both with $\lambda=0.2$. We have plotted the best-fitting models when $\log{(T_\Exp/\yr)}$ is fixed and equal to 4.80 (dashed-blue line), 5.00(dashed-black line) and 5.20 (solid-cyan line). The solid-green line shows the original best-fitting model when the gas expulsion time-scale is also a free parameter. The original values for $\log{(T_\Exp/\yr)}$ is $4.15$ and $3.40$ for MD-RG8.5 and MI-RG8.5 respectively.}
	\label{fig:distributions2}
\end{figure*}

\par We use a least-squares method, to find the model which best matches the observations. We consider the distribution of Galactic GCs from the latest version of \citet{Harris} catalogue in a 2D plane of mass vs radius and split this 2D plane into bins and calculate the normalized frequency of GCs in each bin, i.e. the number of GCs that are in each bin divided by the total number of GCs, so that we have a 2D matrix of these normalized frequencies (hereafter $O$). We make a similar matrix for our simulated clusters (hereafter $S$). We then calculate the sum of the squared residuals between matrices $O$ and $S$
\begin{equation}\label{eq:D}
	D = \sum_{ij} (O_{ij}-S_{ij})^2
\end{equation}

\par By minimizing $D$, we can find the set of initial parameters, given in Table \ref{tab:MCParameters}, which best match the observational distribution. As an additional criterion, we only consider those set of initial parameters for which the distribution of the fraction of SG stars has a sample mean value of $50\pm5\%$. This way we can make sure that the fraction of SG stars in our simulated clusters are consistent with what we see in the observed clusters \citep{DAntona08}. In order to reduce the statistical errors, we do our MC simulations in the neighbourhood of each best-fitting model in the parameter space for 20 different random seed numbers. We then take the mean values of $D$ and the fraction of SG stars as the selection criteria. 

\par Table \ref{tab:MCSimulations} lists our best-fitting models for different tidal fields, concentration of SG stars and dependence of cluster initial radii on the cluster masses. Fig. \ref{fig:distributions} illustrates the outcome of our MC simulations for the following models: (MD-RG2.0, $\lambda=0.1$), (MD-RG4.0, $\lambda=0.2$), (MD-RG8.5, $\lambda=0.1$) and (MI-RG8.5, $\lambda=0.2$) and compares them  with the distribution of the observed clusters. As one can see, our best-fitting models match the distribution of the observed clusters very well. The mass and half-mass radius distributions of our best-fitting models have roughly preserved their initial log-normal distributions and the mean of the distributions have shifted towards lower masses and higher radii respectively which is a direct consequence of gas expulsion. Due to the low number of observed GCs with measured MSP ratios, in our analysis we only fit the mean of the distribution of SG fraction and not the actual shape of the distribution. Fig. \ref{fig:distributions} shows that our best-fitting models have sample means of $50\pm5\%$ for the fraction of SG stars, which is consistent with observations. 

\par According to Table \ref{tab:MCSimulations}, the initial stellar masses of the GCs need to be of order $5-15\times10^5\Msun$ with a gas fraction of at least $\eta=1.0$, meaning that for the significant mass-loss scenario to work we need as much mass in gas as in FG stars. In addition, GCs in stronger tidal fields need to be initially more massive compared to GCs in weaker tidal fields since mass-loss is stronger in stronger tidal fields. Eq. \eqref{eq:massRadius} gives an initial half-mass radius of $\sim1.0\pc$ for our best mass-dependent models roughly equal to that of mass-independent models. The required gas expulsion time-scales are all extremely short, $T_\Exp\sim 10^4$\,yr ($T_{\rm cross}\sim10^5\yr, \tau=0.1$). To see if higher gas expulsion time-scales also lead to acceptable fits, we fixed the value of $\log{(T_\Exp/\yr)}$ to three different values of $4.80$, $5.00$ and $5.20$ respectively
and determine the best fitting-models for the MD-RG8.5 and MI-RG8.5 models with $\lambda=0.20$. Fig. \ref{fig:distributions2} shows that for gas expulsion time-scales larger than $T\geq10^5\yr$, the final properties of clusters are in strong disagreement with the observed clusters, especially for mass-dependent models, implying that the gas expulsion time-scale must have been less than $T=10^5\yr$.

\begin{figure}
	\includegraphics[width=0.45\textwidth]{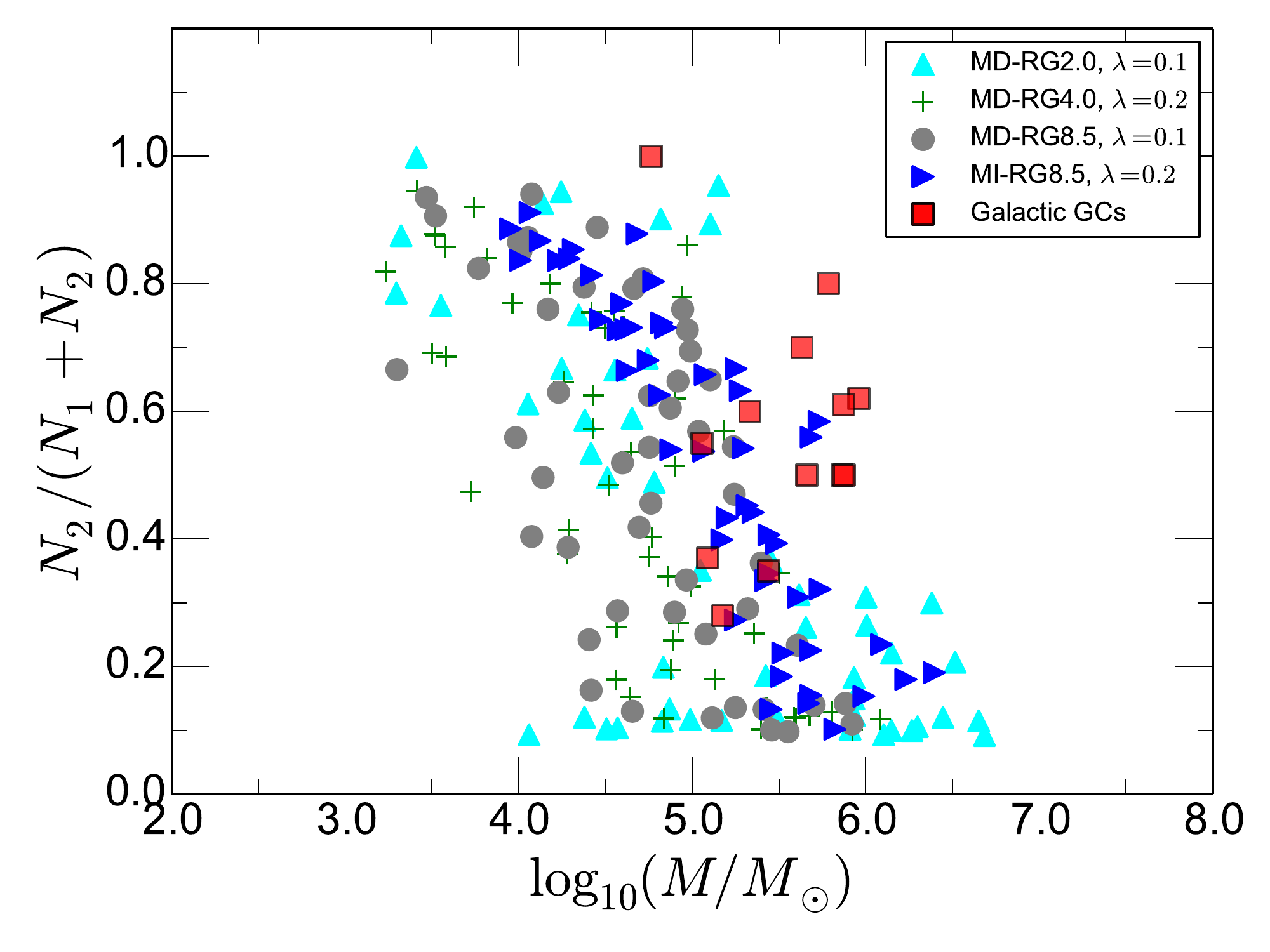}
	\caption{Fraction of SG stars as a function of the cluster mass for the simulated and the observed clusters (filled red squares). Simulated data points show an anti-correlation which is not seen in the observed clusters. In this plot, we show only a fraction of the simulated data points for clarity.} 
	\label{fig:anticorrelation}
\end{figure}

\par Our MC simulations also predict an anti-correlation between the fraction of SG stars and the final mass of GCs as illustrated in Fig. \ref{fig:anticorrelation}. This is due to fact that to increase the fraction of SG stars, GCs need to lose many of their FG stars and since FG stars constitute $\sim 90\%$ of the cluster initial mass, such GCs will have a lower final mass on average.  We have used the Pearson correlation coefficient to quantify this anti-correlation
\begin{equation*}
r_{x,y}=\frac{\Big\langle(x-\langle x\rangle)(y-\langle y\rangle)\Big\rangle}{\sigma_x\sigma_y}
\end{equation*}
where $x$ and $y$ corresponds to $\log_{10}(M_\star/\Msun)$ and $N_2/(N_1+N_2)$ respectively. The penultimate column of Table \ref{tab:MCSimulations} shows the value of the anti-correlation for different models. As one can see this anti-correlation exists in all models, regardless of the tidal field strength or concentration of SG stars. The anti-correlation is more pronounced for mass-independent models with an average $r$ value of $\sim-0.85$ compared to mass-dependent models with $r\sim-0.68$. Observed clusters, denoted by the filled red squares in Fig \ref{fig:anticorrelation}, only exhibit a weak anti-correlation with $r=-0.07$ as the fraction of SG stars is almost independent of the cluster mass. The $90\%$ confidence interval of $r$ for observed GCs, obtained from the bias-corrected and accelerated (BCa) bootstrap method of \citet{Efron}, is [-0.68, 0.61]. As a result the lower confidence bound for the anti-correlation coefficient of observed GCs is marginally in agreement with the mass-dependent models but it is still different from mass-independent models.

\par This discrepancy can be explained in two ways. Either the significant mass-loss scenario does not work or we need to conduct more surveys to find the fraction of different stellar populations for more clusters. In either case, the existence of such an anti-correlation could be used as a diagnostic to test our scenario.

\par The analysis of \citet{Decressin10} on the \citet{Baumgardt07} models show a similar relation between the fraction of SG stars and the number of bound stars. Since the number of bound stars is proportional to the total mass of the cluster, the result of their work matches the anti-correlation that we see in Fig. \ref{fig:anticorrelation}.


\section{Conclusions and Discussion}\label{sec:conclusions}
Using a large grid of \Nbody and Monte Carlo simulations we have studied the consequences of primordial mass-loss for GCs with MSPs to put constraints on their initial conditions and find the best match with observations. We have demonstrated that primordial mass-loss is able to simultaneously reproduce the present-day distribution of GCs in the mass-radius plane (Fig. \ref{fig:distributions}) and explain the large fraction of second generation stars in the cluster. However this is only possible if: (1) the total mass of the gas remaining in the cluster is equal to that of first generation stars $M\sim10^6\Msun$ and (2) very short gas expulsion time-scales of less than $10^5\yr$ which is equal to about one initial crossing time.  In this case, typical initial masses of globular clusters are around $M=10^6\Msun$ and their initial half-mass radii are around $r_h=1\pc$ (Table \ref{tab:MCSimulations}).

\par According to \citet{Decressin10} for a gas cloud with an initial mass of $\sim10^6\Msun$ and an initial half-mass radius of $0.5\pc$, around 50 SNe are needed to unbind the gas cloud. In our case the gas clouds, which are more concentrated than the first generation stars, have initial half-mass radii of $0.1$ and $0.2\pc$. Since the gravitational potential energy scales with $\propto R^{-1}$, our gas clouds will need at least of order $125-250$ SN explosions to be dispersed. According to Fig. 5 of \citet{Decressin10} in a cluster with $M=10^6\Msun$, at most 400 SNe/$\Myr$ will explode within 1 Myr implying that only 40 SNe will go off in $10^5\yr$ which is a factor 3 to 6 below the required limit. As a result SN explosions seem not to able to generate enough energy to expel the gas over the short time-scales we need in our scenario. 

\par In addition, \citet{Krause12} have shown that the Rayleigh-Taylor \citep{Sharp} instability destroys the huge gas shells (superbubbles) made by SN explosions before they build up enough speed to leave the cluster, thus such superbubbles are ineffective in expelling the gas even if they have enough energy. Instead \citet{Krause12} propose accretion onto dark remnants, such as neutron stars and black holes, as a promising mechanism which is capable to overcome the Rayleigh-Taylor instability and lead to rapid gas expulsion. According to their model, the dark remnants become active after the SN phase ($t>35\Myr$) \citep{Krause13} and are able to unbind the intra-cluster medium in $0.03\Myr$ to $0.06\Myr$ depending on if neutron stars also contribute to the gas expulsion \citep{Krause12, Krause13}. 
\par According to \citet{Krause12, Krause13}, This model works for protoclusters whose masses are less than $2\times10^7\Msun$ above which the gas cannot be ejected and will be retained in the cluster. This is intriguing, because firstly such a mass limit encompasses the majority of GCs except for very massive ones such as $\omega$ Cen, which is 10 times more massive, and secondly MSPs in such massive GCs have different iron abundances which could be a result of gas retention \citep{Krause12}. 

\par The gas expulsion time-scales and the initial mass of the gas clouds the we obtain in this work, match results by \citet{Krause12, Krause13} very well. However it is not clear whether GCs can retain gas clouds with masses of order $\sim10^6\Msun$ for $\sim35\Myr$. Observations of young massive star clusters by \citet{Bastian14} and \citet{Hollyhead} show that they have cleared out their natal gas within a few Myrs. If it was the case for GCs as well, then it poses a serious challenge for the scenario proposed by \citet{Krause12, Krause13} and would imply that either stellar winds and supernova explosions do have to expel the gas or that the gas is completely consumed into stars after a few Myr in which case the scenario suggested here would not work.

\par Another possibility is that only the centres of star clusters are gas free, but that the gas is still present in the outer regions. If converted to a physical scale, the half-mass radius of the second generation stars in our best-fitting models is around $\sim0.1\pc$ which is still far less than half-mass radius of the whole cluster ($\sim1\pc$). As a result, it is possible to start with clusters that have a high star formation efficiency and little gas in the very centre, followed by a region with considerable amounts of remaining gas at intermediate radii and then the first generation stars at large radii, and still end up with large numbers of second generation stars after gas expulsion. This should work as long as first and second generation stars are well separated in space, i.e. $a_2/a_1 \ll 1$.

\par In the AGB scenario, second generation stars form after SN explosions or dark remnants have expelled the primordial gas not accreted into stars from the clusters. As a result, the clusters need to accrete significant amounts of unprocessed new gas into their centers to start formation of second generation stars, while at the same time preventing the dark remnants to immediately eject this gas. According to our scenario, when it finally happens, the gas expulsion has to be very rapid. At the moment it is completely unclear if and how this is possible. In addition, accretion of gas is only possible for clusters which move with a low relative velocity to an surrounding gaseous medium \citep{Pflamm}, however second generation stars have been found in almost all massive globular clusters, independent of their orbits and position in the Milky Way. These problems do not exist in the FRMS scenario or any other scenario that form second generation stars within a few Myrs of the first generation ones, since the gas out of which the second generation forms is already in the cluster. 

\par Since the fraction of second generation stars increases as a result of long-term dynamical evolution of the clusters in the galactic tidal field \citep{Decressin08}, the fraction of second generation stars at the end of the gas expulsion phase could be lower than the present-day fraction. This would mean that the gas expulsion time-scales could be larger than what we found here since the fraction of second generation stars at the end of the gas expulsion decreases with the gas expulsion time-scale. However since the lifetimes of most globular clusters are significantly longer than a Hubble time \citep{Baumgardt03}, we do not expect the fraction of second generation stars to change significantly over a Hubble time due to dynamical evolution, and therefore our upper limit of $10^5\yr$ for the gas expulsion time-scale is unlikely to change significantly.

\par The outcome of our simulations shows that fraction of second generation stars is inversely proportional to the final cluster mass (Fig. \ref{fig:anticorrelation}). This anti-correlation, which is in agreement with \citet{Decressin10}, is one of the implications of the primordial mass-loss and can be used to test the feasibility of this scenario. Observations show that such an anti-correlation, albeit weaker, also exist in the Galactic GCs. For our simulated clusters the anti-correlation coefficient ranges from $-0.50$ to $-0.87$, whereas for Galactic GCs it is about $r\sim-0.07$ with a 90\% confidence interval of $[-0.68, 0.61]$. As a result Galactic GCs show a relatively weaker anti-correlation. However, given the 90\% confidence interval on the correlation coefficient of Galactic GCs, the data is also consistent with no anti-correlation or even a positive correlation.

\par The discrepancy between theory and observation might be due to low-number statistics. Also in individual GCs, the number ratio of second generation stars has been measured only over a limited range in radius and observations show that the ratio is varying with radius \citep{Lardo}. Better observations are therefore needed to test if an anti-correlation similar to the one predicted by our models exists in Galactic GCs.

\par In our Monte Carlo simulations we have assumed a log-normal distribution for the initial cluster mass function. Instead of a log-normal relation, one can also assume a power-law distribution ${\rm d}N\propto M^{-\alpha}$ for the cluster mass function, with $\alpha\approx2$, as seen for young massive star clusters in interacting and merging galaxies \citep{Whitmore95, Whitmore99}. \citet{Baumgardt08} for example studied the effect of residual gas expulsion on gas embedded star clusters and found that it is possible to turn a log-normal mass function into a power-law over a Hubble time due to gas expulsion. They found that this effect is almost independent of the strength of the external tidal field or the assumed model for gas expulsion. As a result, our model could also work for an initial power-law distribution. A potential problem for such a mass function could be the over-production of the field halo stars due to the large number of disrupted clusters. A detailed numerical analysis of this effect is beyond the scope of the present paper and can be the subject of a future paper.


\section*{Acknowledgments}
This research was undertaken using Green II GPU supercomputer and other computational resources provided at the Swinburne University of Technology, through the AAL's ASTAC (Astronomy Australia Ltd's Astronomy Supercomputer Time Allocation Committee\footnote{http://www.astronomyaustralia.org.au}) scheme supported by the Australian Government. The authors also would like to thank the anonymous referee for his/her useful comments. PK and HB are members of "Massive star clusters across the Hubble time" team led by Corinne Charbonnel in the International Space Science Institute (ISSI), Bern, Switzerland. PK and HB would like to appreciate ISSI funding and hospitality during the team meetings in Jan 2014 and June 2015 which improved the quality of this work.



\bsp
\label{lastpage}

\end{document}